\documentclass[11pt]{article}

\usepackage[verbose=true,letterpaper,margin=1in]{geometry}

\usepackage[utf8]{inputenc}
\usepackage[T1]{fontenc}
\usepackage{lmodern}
\usepackage{authblk}

\usepackage[hidelinks]{hyperref}
\usepackage{url}

\usepackage{booktabs}
\usepackage{threeparttable}

\usepackage{amsfonts}
\usepackage{amsmath}
\usepackage{amssymb}

\usepackage{nicefrac}
\usepackage{microtype}
\usepackage{graphicx}

\usepackage{natbib}
\setcitestyle{authoryear,aysep={},notesep={, }}

\DeclareMathOperator{\tr}{tr}
\newcommand{\Phirm}{\mathrm{\Phi}}
\newcommand{\ci}{\perp\!\!\!\perp}

\newcommand\blfootnote[1]{%
  \begingroup
  \renewcommand\thefootnote{}\footnote{#1}%
  \addtocounter{footnote}{-1}%
  \endgroup
}

\title{{\textbf{A restricted latent class model with polytomous attributes and respondent-level covariates}}}

\author[1]{Eric Alan Wayman}
\author[2]{Steven Andrew Culpepper}
\author[2]{Jeff Douglas}
\author[1]{Jesse Bowers}
\affil[1]{Independent scholar \protect\\
\texttt{ericwaymanpublications@mailworks.org},  \texttt{bowers.jesse+publications@gmail.com}}
\affil[2]{Department of Statistics, University of Illinois Urbana-Champaign \protect\\
\texttt{sculpepp@illinois.edu}, \texttt{jeffdoug@illinois.edu}}

\date{}

\begin{document}

\maketitle

\begin{abstract}
We present an exploratory restricted latent class model where response data is for a single time point, polytomous, and differing across items, and where latent classes reflect a multi-attribute state where each attribute is ordinal. Our model extends previous work to allow for correlation of the attributes through a multivariate probit specification and to allow for respondent-specific covariates. We demonstrate that the model recovers parameters well in a variety of realistic scenarios, and apply the model to the analysis of a particular dataset designed to diagnose depression. The application demonstrates the utility of the model in identifying the latent structure of depression beyond single-factor approaches which have been used in the past.
\end{abstract}

\section{Introduction}

\blfootnote{The version of record of this article, first published in \emph{Behaviormetrika}, is available online at the publisher's website: \url{https://doi.org/10.1007/s41237-025-00271-8}}

Latent class models treat respondents as coming from a finite number of unobservable or latent groups. In their most general form they provide an exploratory method for clustering respondents but do not admit the most convenient interpretation. A more parsimonious and interpretable alternative to latent class models are restricted latent class models, or RLCMs \citep{haertel1984application, haertel1990continuous}, where the restrictions allow researchers to uncover structure by imposing constraints on certain parameters. These restrictions often allow an interpretation of the class or latent structure in terms of the relationship of that latent structure to the response variables. The latent structure in many types of latent variable models is assumed to be multidimensional: in latent trait models (such as factor analysis) the values taken by each dimension are real numbers, whereas in RLCMs the values taken are from subsets of positive integers and are thus discrete (each dimension of the latent state in these models is referred to as an attribute). In medicine and health quality of life, such discrete values are desirable for the purpose of diagnostic classification, and the results of the model estimation can help practitioners select an appropriate treatment. However, in the past, instruments for the purpose of such diagnosis were often developed with a latent trait framework in mind.

Although RLCMs have been applied in multiple fields, such as education \citep{torre2004higher, chen2020sparse}, organizational behavior \citep{sorrel2016validity}, and psychological assessment \citep{chen2018introducing, templin2006measurement, chen2015statistical, torre2011generalized}, they have not yet seen significant usage for diagnosis in the areas of medicine and mental health. One reason is that most RLCMs have only binary rather than polytomous attributes, which strictly limits the characterization of the latent state of the condition. RLCMs with polytomous attributes include \citet{chen2013general} (different levels possible for each dimension, confirmatory model), \citet{sun2013polytomous} (same levels for each dimension, confirmatory model), and \cite{he2023sparse} (same levels for each dimension, exploratory model).

Another factor limiting the applicability of RLCMs thus far is that most models do not relate respondent-specific covariates to the respondent's latent state, which practitioners might use to help their decision-making (for instance, none of the aforementioned three models for polytomous attributes allow for covariates). One use of covariates is to measure the intervention effect of a particular treatment, i.e. to see whether or not a particular intervention contribute to differences in classification. Another is that additional information about respondents can be used to help improve classification.

There are identifiability results involving covariates: \citet{ouyang2022identifiability} proved identifiability of an RLCM with respondent-specific covariates and binary attributes. \citet{yigit2023extending} present a model with binary attributes, where respondent-specific covariates are related to both item response probabilities and the respondent's latent state.

The model presented in this manuscript is for polytomous attributes, and relates covariates to the respondent's latent state through the use of a multivariate probit specification. This use of the multivariate probit differs from previous approaches, which used a Dirichlet distribution with binary attributes \cite[e.g., see][]{chen2023bayesian, liu2023identifiability}, higher-order factor models \cite[]{culpepper2023inferring,torre2004higher} (both with binary attributes), a higher-order model with polytomous attributes but for binary rather than polytomous data \citep{ma2021higher}, or a multivariate probit specification for the case of binary attributes only \citep{templin2008robustness}.

The two closest models to this one are those of \citet{culpepper2023inferring}, which has a simpler structure describing the binary attributes and has no covariates, and \citet{he2023sparse}, which has a general Dirichlet prior for the polytomous attributes and does not account for covariates. Comparing our model to these, we note the following: (1) our model and its accompanying software incorporate covariates where these others do not, and (2) our model is a more parsimonious way of representing the relationship amongst the polytomous attributes than the full Dirichlet type of prior, but is more general than the higher order factor model, which imposes partial structure on the correlation matrix, as opposed to the unstructured correlation matrix of our multivariate probit specification.

This paper makes the following contributions to the RLCM literature. First, the model we present allows for respondent-specific covariates in an RLCM with polytomous attributes. Second, our model uses a multivariate probit specification to model correlation between the latent polytomous attributes. Third, we introduce a particular prior for latent attribute thresholds which allows for Markov Chain Monte Carlo (MCMC) sampling in the polytomous attribute case where for some replications membership in the top level may be empty. Fourth, we demonstrate the use of parameter expansion for sampling the parameter expressing association between the attributes and covariates, the thresholds which divide the categories, and the underlying attribute polychoric correlation matrix. Fifth, we present a model selection procedure which uses posterior predictive checks \citep{gelman1996posterior, crespi2009bayesian} to select the most parsimonious model for which the data-generating process results in the salient features of the observed data being reproduced. Sixth, we demonstrate the applicability of this framework to the mental health setting. 

The structure of our paper is as follows. In the next section, we describe the major components of our model, namely the measurement model, structural model, and a monotonicity condition. In the Bayesian Model section, we describe the the conditional independence relationships between the variables and the assumed relationships between them (likelihood and priors). In the Algorithm section, we show the parameter expansion we perform in order to obtain a posterior from which we can easily sample, and explain the MCMC sampling algorithm. In the Simulation Studies section, we show the results of two simulation studies that demonstrate the model's efficacy in a variety of scenarios. In the Application section, we apply the model to response data from a questionnaire designed for the diagnosis of depression to show our model's efficacy in a real-world diagnostic situation. We close with a discussion.

\section{Overall approach}

Our approach is designed to estimate the parameters of two models: the ``measurement model,'' which relates a respondent's observed data to that respondent's latent state, and the ``structural model,'' which relates a respondent's to the respondent-specific covariates. We also describe a monotonicity condition which relates the latent state to the response values.

\subsection{Measurement model}

In what follows, for \(Q \in \mathbb{N}\), let \([Q]\) denote the set \(\{1, 2, \ldots, Q\}\). We have \(N\) respondents, each of whom responds to a questionnaire of \(J\) items. Each item \(j \in [J]\) consists of \(M_j\) possible responses, numbered from \(0\) to \(M_j - 1\). Each respondent \(n\) provides a vector of responses \(Y_n = (Y_{n1}, \ldots, Y_{nJ})\), where for all \(n \in [N]\), for all \(j \in [J]\), we have \(Y_{nj} \in \left\{0, \ldots, M_j - 1\right\}\). We also assume that we observe for each respondent a vector of covariates represented as a row vector: for all \(n \in [N]\), \(X_n \in \mathbb{R}^{1 \times D}\), where \(D\) is the number of covariates plus one since an intercept is included.

We assume each respondent \(n\) has a latent state \begin{equation}
  \alpha_n = (\alpha_{n1}, \ldots, \alpha_{nK}) \in \left\{0, \ldots, L - 1\right\}^K
\end{equation}

\noindent of \(K\) ordinal attributes, arranged as a row vector, where for all \(k \in [K]\) the number of possible levels attribute \(k\) can take is \(L\), a natural number. We define \(A_L = \left\{0, \ldots, L - 1\right\}^K\), the set of all possible latent state vectors.

The measurement model is a cumulative probit model \citep[pages 211-212]{agresti2015foundations}: for all \(n \in [N]\), \(j \in [J]\), and \(m \in \left\{0, \ldots, M_j - 1\right\}\) \begin{equation}
  \Phirm^{-1}[P(Y_{nj} \leq m \mid \alpha_n, \beta_j, \kappa_j)] = \kappa_{j, m + 1} - d_n \beta_{j}
\end{equation}

\noindent where \(\Phirm\) is the cdf of the standard normal distribution, \(\kappa_{j0} < \kappa_{j1} < \cdots < \kappa_{j M_j}\) with \(\kappa_{j0} = -\infty\), \(\kappa_{j1} = 0\), \(\kappa_{j Mj} = \infty\) for identifiability purposes, and \(\beta_j\) are regression coefficients or item parameters relating item \(j\) to the respondent's attribute profile \(d_n\).

The row vector \(d_n\) is a ``design vector'' that codes the main effects and selected interactions corresponding to the respondent's latent state. The design vector uses a ``cumulative coding'' \citep[page 417]{he2023sparse} defined as follows. For an arbitrary latent state vector \(\alpha_n = (\alpha_{n1}, \ldots, \alpha_{nK})\), we define, for all \(k \in [K]\), the functions \(d_L^k: A_L \to \{0, 1\}^L\) by \(d_L^k(\alpha_n) = (I(\alpha_{nk} \geq 0), I(\alpha_{nk} \geq 1), \ldots, I(\alpha_{nk} \geq L - 1))\). We define the function \begin{equation}
  d_L: A_L \to \{0, 1\}^{\prod_{i=1}^K L}
\end{equation}

\noindent by \(d_L(\alpha_n) = d_L^1(\alpha_n) \otimes d_L^2(\alpha_n) \otimes \cdots \otimes d_L^K(\alpha_n)\). We allow for practioners to choose the ``order'' of the model, which refers to the maximum number of attributes allowed in an effect; we denote the resulting dimension of the design vector \(d_n\) as \(H\).

\subsection{Monotonicity condition}

We say that for two arbitrary \(\alpha_n\)'s, say \(\alpha_{n_1}\) and \(\alpha_{n_2}\), \(\alpha_{n_1} = (\alpha_{n_1 1}, \ldots, \alpha_{n_1 K})\) and \(\alpha_{n_2} = (\alpha_{n_2 1}, \ldots, \allowbreak\alpha_{n_2 K})\), that \(\alpha_{n_1} \geq \alpha_{n_2}\) iff for all \(k \in [K]\) we have \(\alpha_{n_1 k} \geq \alpha_{n_2 k}\).

We assume that the latent state \(\alpha_n\) is an ordinal random variable in the sense that for any item, for any response level, the probability of responding with a higher response level is increasing in \(\alpha_n\). This assumption, a monotonicity condition, is from \citet[page 928]{culpepper2019exploratory}; in mathematical terms it is: for all \(n \in [N]\) and \(j \in [J]\), \begin{equation}\label{monotoncond}
\forall\, u, v \in A_L \quad u \geq v \implies p(Y_{nj} > m \mid u, \beta_j, \kappa_j) \geq p(Y_{nj} > m \mid v, \beta_j, \kappa_j).
\end{equation}

\noindent The above constraint equivalently is \(\forall\, u, v \in A_L \quad u \geq v \implies d_u \beta_j \geq d_v \beta_j\), where \(d_u\) is the design vector associated with \(u\). We note that this is a stricter condition than that which was used in \citet{chen2020sparse} and \citet{he2023sparse}.

\subsection{Structural model}

We model \(\alpha_n\) as arising from a discretized version of a continuous multivariate normal \(\alpha^\ast_n\) \citep{mcdonald1967nonlinear,ashford1970multi,christoffersson1975factor,muthen1978contributions}. This multivariate probit model is characterized by: \begin{equation}\label{mvp}
p(\alpha_n \mid \gamma, \lambda, R) = \int_{\gamma_{K, \alpha_{nK}}}^{\gamma_{K, \alpha_{nK} + 1}} \ldots \int_{\gamma_{1, \alpha_{n1}}}^{\gamma_{1, \alpha_{n1} + 1}} \phi_K(\alpha_n^{\ast}; X_n \lambda, R) d\alpha_n^{\ast}
\end{equation}

\noindent where \(\phi_K(x; A, B)\) denotes the density of a multivariate normal with number of attributes \(K\), variable \(x\) (a row vector), mean \(A\), and covariance \(B\), \(R\) is a positive definite correlation matrix, and for all \(k \in [K]\), \(\gamma_{k0} < \gamma_{k1} < \cdots < \gamma_{k L}\) with \(\gamma_{k0} = -\infty\), \(\gamma_{k1} = 0\), \(\gamma_{k L} = \infty\) set for identifiability purposes. \(\lambda\) is the matrix of coeffients which expresses how the covariates relate to the latent state.

Note that we choose a correlation structure rather than a covariance structure due to identifiability reasons \citep{lawrence2008bayesian}. We treat \(\alpha_n\) and \(\alpha_n^\ast\) as row vectors. Also, for all \(k \in [K]\) we treat \(\gamma_k\) as a column vector in the extended reals with \(L + 1\) entries, \(L + 1 - 3 = L - 2\) of which are free parameters, and denote the matrix \(\gamma = (\gamma_1, \ldots, \gamma_K)\).

\section{Bayesian model}

\begin{figure*}
  \centering
  \includegraphics[width=0.5\linewidth]{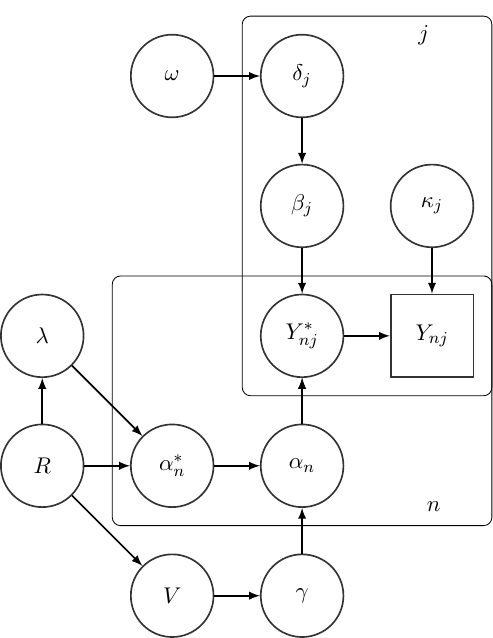}
  \caption{\label{fig:dgm} Directed graphical model (for alt text, see Supplementary Material D)}
\end{figure*}

In order to build a model from which we can sample using a multiple-block Metropolis-Hastings algorithm \citep{chib2011introduction}, we formulate our Bayesian model as a directed graphical model \citep[pages 310-311]{murphy2012machine}, whose graph \(G\) is displayed in Figure \ref{fig:dgm} using plate notation \citep[pages 320-322]{murphy2012machine}.

\begin{table}
\small
\begin{center}
\caption{\label{tab:paramslist} Parameter descriptions}
\vspace{0.5\baselineskip}
\begin{tabular}{ll}
\toprule
Parameter & Description \\
\midrule
\(Y\) & responses \\
\(\alpha\) & latent states \\
\(Y^{\ast}\) & augmented variables for responses \\
\(\alpha^{\ast}\) & augmented variables for latent states \\
\(\kappa\) & thresholds for measurement model \\
\(\beta\) & slope parameters relating latent states to responses \\
\(\delta\) & activation indicator variables for \(\beta\) \\
\(\omega\) & part of prior for \(\delta\) \\
\(\gamma\) & thresholds for multivariate probit specification \\
\(\lambda\) & slope parameters relating covariates to latent states \\
\(R\) & polychoric correlation matrix for latent states \\
\(V\) & diagonal of covariance matrix \\
\bottomrule
\end{tabular}
\end{center}
\end{table}

We write \(Z = (Y, \alpha, \alpha^\ast, Y^\ast, \theta)\), where \(\theta = (\kappa, \beta, \delta, \omega, \gamma, \lambda, R, V)\) and where each variable with an asterisk as a superscript is an auxiliary variable introduced for computational purposes. Table \ref{tab:paramslist} describes the variables. As a directed graphical model, \(p(Z)\) admits a recursive factorization according to \(G\) \cite[page 46]{lauritzen1996graphical}, namely that \(p(Z) = \prod_{z \in Z} p(z \mid \text{pa}(z))\) where \(\text{pa}(z)\) indicates the parents of vertex \(z\). We observe that by the recursive factorization, we have the following representation: \begin{align}
  p(Z) & = p(Y \mid Y^\ast, \kappa) \cdot p(Y^\ast \mid \beta, \alpha) \cdot p(\kappa) \cdot p(\beta \mid \delta) \cdot p(\delta \mid \omega) \cdot p(\omega) \notag\\
& \quad\quad \cdot p(\alpha \mid \alpha^\ast, \gamma) \cdot p(\gamma \mid V) \cdot p(R, V) \cdot p(\alpha^\ast \mid R, \lambda) \cdot p(\lambda \mid R) \label{originalmodel}
\end{align}

\noindent where we have taken an additional step, writing \(p(R, V) = p(V \mid R) \cdot p(R)\). In the above expression, we have amalgamated across subscripts; for example, \(\kappa\) is the set of all \(\kappa_j\). We refer to \eqref{originalmodel} as the ``original model.''

Recall that the model \(p(Z)\) admits a recursive factorization according to \(G\) if and only if the model obeys the directed local Markov property relative to \(G\), namely that for any variable \(z \in Z\) we have \(z \ci \text{nd}(z) \mid \text{pa}(z)\), where \(\text{nd}(z)\) refers to the non-descendants of \(z\) not including \(\text{pa}(z)\) \citep[page 51, Theorem 3.27]{lauritzen1996graphical}.

In the following we specify the forms of assumed relationships (likelihood and priors) which appear in the above factorization.

\subsection{Likelihood and prior related to observed data}

We implement the data augmentation \citep{tanner2010data} method of \citet{albert1993bayesian}, namely, we assume for all \(n \in [N]\) and \(j \in [J]\) that \begin{equation}
  p(Y_{nj}^{\ast} \mid d_n, \beta_j) = \phi(Y_{nj}^\ast;\, d_n \beta_j, 1)
\end{equation}

\noindent where \(\phi(x; a, b)\) denotes the pdf of the normal distribution with variable \(x\), mean \(a\), and variance \(b\), as well as \begin{equation}
  p(Y_{nj} \mid Y_{nj}^\ast, \kappa_j) = \sum_{m=0}^{M_j - 1} I(Y_{nj} = m) \cdot I\left(Y_{nj}^\ast \in (\kappa_{jm}, \kappa_{j, m + 1}]\right)
\end{equation}

\noindent which yields \begin{equation}\label{itemresponseprob}
  p(Y_{nj} \mid \alpha_n, \beta_j, \kappa_j) = \int_{\kappa_{j Y_{nj}}}^{\kappa_{j(Y_{nj} + 1)}} \phi(Y_{nj}^\ast;\, d_n \beta_j, 1) dY_{nj}^\ast.
\end{equation}

We stipulate \(p(\kappa_j) = I(-\infty = \kappa_{j0} < 0 = \kappa_{j1} < \cdots < \kappa_{j M_j} = \infty)\).

\subsection{Priors related to coefficients for measurement model}

Similar to \cite{chen2020sparse}, we implement variable selection for \(\beta\) using the method of \citet{kuo1998variable}: for each \(j \in [J]\) we use a vector \(\delta_j \in \{0, 1\}^H\) to indicate for each \(h \in [H]\) (i.e. each component of \(\beta_j\)) whether \(\beta_{hj}\) should be equal to zero or have a normal distribution (a spike-slab prior with a Dirac spike and normal slab). Since the values for each \(\beta_j\) are affected by the monotonicity condition \eqref{monotoncond}, we include an indicator function in the prior \(p(\beta_j \mid \delta_j)\). Specifically, we assume \citep{malsiner2011comparing} \begin{equation}
  p(\beta_j \mid \delta_j) = c_j(\delta_j) \cdot I(\beta_j \in \mathcal{R}_j) \cdot \prod_{h=1}^H p(\beta_{hj} \mid \delta_{hj}), \quad p(\delta_j \mid \omega) = \prod_{h=1}^H p(\delta_{hj} \mid \omega)\end{equation} \begin{equation}p(\beta_{hj} \mid \delta_{j}) = p(\beta_{hj} \mid \delta_{hj}) = I(\delta_{hj} = 0) \cdot \mathrm{\Delta}(\beta_{hj}) + I(\delta_{hj} = 1) \cdot \phi(\beta_{hj}; 0, \sigma_{\beta}^2)
\end{equation}

\noindent where \(\mathrm{\Delta}\) denotes the Dirac delta generalized function, \(\delta_{hj} \mid \omega \sim \text{Bernoulli}(\omega)\), where \(\sigma_{\beta}^2\) is a hyperparameter, and where the indicator function requires \(\beta_j\) fall into a space \begin{equation}
\mathcal{R}_j := \Big\{\beta_j \in \mathbb{R}^H : \forall\, u, v \in A_L \quad u \geq v \implies d_u \beta_j \geq d_v \beta_j \Big\}
\end{equation}

\noindent namely the space for which \(\beta_j\) satisfies the monotonicity constraint \eqref{monotoncond}.

As in \citet[page 132]{chen2020sparse}, we let \(\omega \sim \text{Beta}(\omega_0, \omega_1)\), where \(\omega_0, \omega_1\) are hyperparameters.

\subsection{Priors for structural model}

We use data augmentation for the latent state, assuming for all \(n \in [N]\) \begin{equation}
  \alpha_n^\ast \mid \lambda, R \sim N_K(X_n \lambda, R)
\end{equation}

\noindent a multivariate normal with mean \(X_n \lambda\) and covariance matrix \(R\) (where \(R\) is restricted to be a correlation matrix) and also \begin{equation}
  p(\alpha_n \mid \alpha_n^\ast, \gamma) = \prod_{k=1}^K I(\alpha_{nk}^\ast \in (\gamma_{k \alpha_{nk}}, \gamma_{k(\alpha_{nk} + 1)}])
\end{equation}

\noindent yielding \begin{equation}
  p(\alpha_n \mid \gamma, \lambda, R) = \int_{\gamma_{K\alpha_{nK}}}^{\gamma_{K(\alpha_{nK} + 1)}} \ldots \int_{\gamma_{1\alpha_{n1}}}^{\gamma_{1(\alpha_{n1} + 1)}} \phi_K(\alpha_n^{\ast}; X_n \lambda, R) d\alpha_n^{\ast}.
\end{equation}

\noindent We let \(\lambda \mid R \sim N_{D, K}(0, I_D \otimes R)\).

We factor the prior for the \(\gamma_k\), \(R\), and \(V\) as follows: making use of the directed local Markov property, we have that \(\gamma \ci R \mid V\), so we have that \begin{equation}\label{gammaRV}
  p(\gamma, R, V) = p(\gamma \mid R, V) \cdot p(R, V) = p(\gamma \mid V) \cdot p(R, V).
\end{equation}

\noindent Regarding the vectors of thresholds \(\gamma_k\): for each level \(l \in \{2, 3, \ldots, L - 1\}\) we assume \(\gamma_{kl} \ci \gamma_{k,l-2}, \gamma_{k, l-3}, \ldots, \gamma_{k3}, \gamma_{k2} \mid \gamma_{k, l-1}, v_k\), yielding \begin{align}
  p(\gamma_k \mid v_k) & = p(\gamma_{k, L-1} \mid \gamma_{k, L-2}, v_k) \cdot p(\gamma_{k, L-2} \mid \gamma_{k, L-3}, v_k) \notag\\
  & \quad\quad \cdot \cdots \cdot p(\gamma_{k3} \mid \gamma_{k2}, v_k) \cdot p(\gamma_{k2} \mid v_k) \label{thresholdprior}
\end{align}

\noindent and introduce a new prior for \(\gamma\). This new prior is a left-truncated exponential prior for each threshold, where the constant part of the rate parameter, namely \(a\), should be chosen so that the prior density is relatively diffuse: \begin{align}
  & p(\gamma_{kl} \mid \gamma_{k,l-1}, v_k) \notag\\
  & = I\left(\gamma_{kl} \in (\gamma_{k, l-1}, \infty)\right) \cdot a v_k^{1/2} \cdot \exp\left[-a v_k^{1/2} \cdot (\gamma_{kl} - \gamma_{k,l-1})\right].
\end{align}

\noindent The above prior works out such that the full conditional distributions from which we sample for each level are uniform except for the highest level, the full conditional for which is a left-truncated exponential (see Supplementary Material B for details). If a uniform was assumed rather than a left-truncated exponential for each threshold, the full conditional for the rightmost threshold less than \(\infty\) would be a uniform distribution, but when sampling, there may be iterations for which there is no respondent whose latent state falls into the top bin, in which case the range of the uniform would be infinite and thus the distribution would not be properly defined. In contrast, our prior ensures that even in such a case the sampling algorithm is still computationally tractable since unlike a uniform distribution, a left-truncated exponential can accomodate a support which is unbounded on the right.

We deploy a prior for the correlation matrix similar to \citet{barnard2000modeling}. Specifically, we decompose a positive definite symmetric \(\Sigma\) as \(V^{1/2} R V^{1/2}\) where \(V\) is of the form \(V = \text{diag}(v_1, \ldots, v_K) \in \mathbb{R}^{K \times K}\) and for all \(k \in [K]\), \(v_k > 0\) and the joint density for \(R\) and \(V\) is \begin{equation}
  p(R, V) \propto (\det{R})^{-\frac{1}{2}(v_0 + K + 1)} \cdot \prod_{k \in [K]} \left[ \exp\left(-\frac{1}{2} v_k^{-1} A_{kk}\right) \cdot v_k^{-\frac{1}{2}(v_0 + 2)}\right]
  \end{equation}

\noindent where the \(A_{kk}\) are the diagonal elements of \(R^{-1}\) and where \(v_0 > K - 1\) is a hyperparameter.

\subsection{Integrating out the variance parameter}

We are interested in estimating \(R\) but not \(V\): we note that \(V\) can be eliminated from the original model using \eqref{gammaRV} and taking the integral of \eqref{originalmodel} with respect to \(V\). This yields \begin{align}
p(Z \setminus V) & = p(Y \mid Y^\ast, \kappa) \cdot p(Y^\ast \mid \beta, \alpha) \cdot p(\kappa) \cdot p(\beta \mid \delta) \cdot p(\delta \mid \omega) \cdot p(\omega) \notag\\
& \quad \cdot p(\alpha \mid \alpha^\ast, \gamma) \cdot p(\gamma, R) \cdot p(\alpha^\ast \mid R, \lambda) \cdot p(\lambda \mid R).
\end{align}

\subsection{Parameter expansion}

Since the original model is not easy to sample from directly using a multiple-block Metropolis-Hastings algorithm, we make use of a technique similar to parameter expansion \citep{liu1999parameter} or conditional augmentation \citep{meng1999seeking}, performing transformations similar to those in \citet{lawrence2008bayesian} and \citet{zhang2020parameter} to produce a model from which we can easily sample. We refer to that model as the ``transformed model.'' Our Metropolis-within-Gibbs algorithm \citep{chib2011introduction} samples from the transformed model and at each step performs the inverse of the transformation, resulting in a sample from the original model.

We transform \(Z\) to \(\widetilde{Z}\) using the change of variables formula \(p_{\widetilde{Z}}(\widetilde{z}) = p_{Z}(g^{-1}(\widetilde{z})) \cdot |\det{J_{g^{-1}}(\widetilde{z})}|\), where \(g\) takes values of \(Z\) to values of \(\widetilde{Z}\) and \(J_{g^{-1}}(\widetilde{z})\) is the Jacobian matrix of \(g^{-1}\) evaluated at \(\widetilde{z}\). The transformation is \(\widetilde{\alpha^\ast} = \alpha^\ast V^{1/2}\), \(\widetilde{\gamma} = \gamma V^{1/2}\), \(\widetilde{\lambda} = \lambda V^{1/2}\), and \(\Sigma\ = V^{1/2} R V^{1/2}\). See Supplementary Material A for details.

\section{Algorithm}

We use a Metropolis-within-Gibbs algorithm to approximate the posterior: we sequentially update values of parameters. Derivations of the results for \((\alpha_{nk}, \widetilde{\alpha^\ast}_{nk})\), for \(\widetilde{\gamma}_{kl}\), for \((\Sigma, \widetilde{\lambda})\), and for \(\omega\) are shown in Supplementary Material B.

\subsection{Augmented data and measurement thresholds}

Denoting \(Y_j^{\ast} = (Y_{1j}^{\ast}, \ldots, Y_{nj}^{\ast})^\prime\), for each \(j \in [J]\) we sample \((\kappa_j, Y_j^{\ast})\) using the method of \citet{cowles1996accelerating}; this is a Metropolis step which relies on a proposal density. Note that careful attention needs to be dedicated to choosing a value for the proposal standard deviation \(\sigma_{\kappa}^2\).

\subsection{Beta and delta}

The strategy for updating \(\beta_{hj}\) and \(\delta_{hj}\) extends the binary case of \citet{chen2020sparse} to the setting of polytomous attributes with polytomous responses. The lower bound for \(\beta_{hj}\) when \(\delta_{hj} = 1\), \(L_{hj}\), differs from that of \citet{chen2020sparse} as shown in \eqref{monotoncondcase2} (this equation appears as the result of a derivation, shown in the Appendix).

Our Gibbs step for \(\delta_{hj}\) is collapsed on \(\beta_{hj}\), namely it is \(\delta_{hj} \mid \alpha, \beta_{(h)j}, Y_j^\ast, \omega\), a Bernoulli where \begin{align}\label{deltaprob}
& p\left(\delta_{hj} = 1 \mid \alpha, \beta_{(h)j}, Y_j^\ast, \omega\right) = \Bigg[(1 - \omega) + \omega \cdot \left(\Phirm\left(\frac{-L_{hj}}{\sigma_{\beta}}\right)\right)^{-1} \cdot \left(\frac{c_2^2}{\sigma_{\beta}^2}\right)^{1/2} \notag\\
    & \quad\quad \cdot \exp\left(\frac{c_1^2}{2c_2^2}\right) \cdot \Phirm\left(\frac{-(L_{hj} - c_1)}{c_2}\right)\Bigg]^{-1} \notag\\
& \quad\quad \cdot \omega \cdot \left(\Phirm\left(\frac{-L_{hj}}{\sigma_{\beta}}\right)\right)^{-1} \cdot \left(\frac{c_2^2}{\sigma_{\beta}^2}\right)^{1/2} \cdot \exp\left(\frac{c_1^2}{2c_2^2}\right) \cdot \Phirm\left(\frac{-(L_{hj} - c_1)}{c_2}\right).
\end{align}

\noindent In \eqref{deltaprob}, \begin{equation*}
  c_2^2 = \left[\left(d^{\, \prime} d\right)_{hh} + \frac{1}{\sigma_{\beta}^2}\right]^{-1}
\end{equation*}

\noindent where \(\left(d^{\, \prime} d\right)_{hh}\) refers to the entry in row \(h\) and column \(h\) of the \(H \times H\) matrix \(d^{\, \prime} d\). Also in \eqref{deltaprob}, \begin{equation*}
  c_1 = c_2^2 \cdot \left(d^{\, \prime} Y_j^{\ast} - \left(d^{\, \prime} d\right)_{(h)} \beta_{j(h)}\right)_h
\end{equation*}

\noindent is the entry in row \(h\) of the \(H \times 1\) vector resulting from the calculation, where \(\left(d^{\, \prime} d\right)_{(h)}\) refers to \(d^{\, \prime} d\) with column \(h\) eliminated and where \(\beta_{j(h)}\) refers to the column vector \(\beta_j\) with element \(h\) eliminated.

\(\beta_{hj}\) is sampled from its full conditional, namely \(p(\beta_{hj} \mid \delta_j, Y_j^\ast, \alpha)\), which is a point mass at \(\beta_{hj} = 0\) when \(\delta_{hj} = 0\), and when \(\delta_{hj} = 1\), we have \begin{equation}
  p(\beta_{hj} \mid \delta_{hj}, Y_j^\ast, \alpha) = I\left(\beta_{hj} \in (L_{hj}, \infty)\right) \frac{\phi(\beta_{hj}; c_1, c_2^2)}{[1 - \Phirm(L_{hj}; c_1, c_2^2)]}
\end{equation}

\noindent a left-truncated normal whose left-truncation point is \(L_{hj}\) and whose underlying mean and variance are \(c_1\) and \(c_2^2\) respectively (when \(h = 0\), \(L_{hj} = -\infty\) and the density is that of a normal distribution).

\subsection{Latent states and related auxiliary variables}

For each \(n \in [N]\) and \(k \in [K]\), when sampling \(\alpha_{nk}\) our Gibbs step collapses on \(\widetilde{\alpha^\ast}_{nk}\): we sample \(\alpha_{nk}\) from \(p(\alpha_{nk} \mid \alpha_{n(k)}, \gamma_k, Y_{nj}^\ast, \beta_j)\), a categorical distribution with probability \((\sum_{l=0}^{L - 1} p_l)^{-1} \cdot p_l\) for each \(l \in \{0, 1, \ldots, L - 1\}\) where \begin{equation*}
  p_l = \left[\prod_{j=1}^J \phi(Y_{nj}^\ast; d_n \beta_j, 1)\right] \cdot \left[\Phirm\left(\frac{\widetilde{\gamma}_{k, l + 1} - \mu_{nk}}{\sigma_k}\right) - \Phirm\left(\frac{\widetilde{\gamma}_{k l} - \mu_{nk}}{\sigma_k}\right)\right]
\end{equation*}

\noindent with \begin{equation}
  \mu_{nk} = x_n \lambda_k + (\widetilde{\alpha^{\ast}}_{n(k)} - x_n \lambda_{(k)})\Sigma_{(k)(k)}^{-1} \Sigma_{(k)k}
\end{equation}

\noindent and \begin{equation}
  \sigma_k^2 = \Sigma_{kk} - \Sigma_{k(k)} \Sigma_{(k)(k)}^{-1} \Sigma_{(k)k}.
\end{equation}

\noindent We then use that value to sample \(\widetilde{\alpha^{\ast}}_{nk}\) from its full conditional, \begin{align}
  & p(\widetilde{\alpha^{\ast}}_{nk} \mid \widetilde{\lambda}, \Sigma, \alpha_{nk}, \widetilde{\gamma}_k) \notag\\
  & = I(\widetilde{\alpha^{\ast}}_{nk} \in (\widetilde{\gamma}_{k\alpha_{nk}}, \widetilde{\gamma}_{k,{\alpha_{nk} + 1}})) \frac{\phi(\widetilde{\alpha^{\ast}}_{nk}; \mu_{nk}, \sigma_k^2)}{\Phirm(\widetilde{\gamma}_{k,\alpha_{nk} + 1}; \mu_{nk}, \sigma_k^2) - \Phirm(\widetilde{\gamma}_{k\alpha_{nk}}; \mu_{nk}, \sigma_k^2)}
\end{align}

\noindent which is a truncated normal density with left and right truncation points \(\widetilde{\gamma}_{k\alpha_{nk}}\) and \(\widetilde{\gamma}_{k,\alpha_{nk} + 1}\) respectively and whose underlying normal random variable has mean \(\mu_{nk}\) and variance \(\sigma_k^2\).

\subsection{Thresholds for latent state levels}

The full conditional for \(\widetilde{\gamma}_{kl}\) for \(k \in [K]\) depends on the value for \(l\). For \(l \in \{2, 3, \ldots, L - 2\}\), we sample \(\widetilde{\gamma}_{kl}\) from its full conditional \(p(\widetilde{\gamma}_{kl} \mid \widetilde{\gamma}_{k,l-1}, \widetilde{\gamma}_{k,l+1}, \widetilde{\alpha^\ast})\) which is a continuous uniform distribution on the range \begin{equation}
  \left(\max\left(\max_{n \in [N]:\, \alpha_{nk} = l - 1} \left(\widetilde{\alpha^{\ast}}_{nk}\right), \widetilde{\gamma}_{k,l-1}\right), \min\left(\min_{n \in [N]:\, \alpha_{nk} = l} \left(\widetilde{\alpha^{\ast}}_{nk}\right), \widetilde{\gamma}_{k,l+1}\right)\right).
\end{equation}

\noindent For \(l = L - 1\), the full conditional for \(\widetilde{\gamma}_{kl}\) is a truncated exponential density, \begin{align}
  p(\widetilde{\gamma}_{kl} \mid \widetilde{\gamma}_{k,l-1}, \widetilde{\alpha^\ast}) & = c \cdot I\left(\widetilde{\gamma}_{kl} \geq \max\left(\max_{n \in [N]:\, \alpha_{nk} = l - 1} \left(\widetilde{\alpha^{\ast}}_{nk}\right), \widetilde{\gamma}_{k, l-1}\right)\right) \notag\\
  & \quad\quad \cdot I\left(\widetilde{\gamma}_{kl} < \min\left(\min_{n \in [N]:\, \alpha_{nk} = l} \left(\widetilde{\alpha^{\ast}}_{nk}\right), \infty\right)\right) \notag\\
  & \quad\quad \cdot \exp\left(-a \widetilde{\gamma}_{kl}\right).
\end{align}

\subsection{Covariance matrix and slope parameter for covariates}

Regarding \((\widetilde{\lambda}, \Sigma)\), our first Gibbs step samples \(\Sigma\) collapsed on \(\widetilde{\lambda}\) from \begin{equation}
  p(\Sigma \mid \widetilde{\alpha^\ast}) = c \cdot (\det{\Sigma})^{-([K + 1 + N] + K + 1)/2} \cdot \text{etr}\left\{-\frac{1}{2}(I_K + S) \Sigma^{-1}\right\}
\end{equation}

\noindent an inverse Wishart with matrix parameter \(I_K + S\) and scalar parameter \(K + 1 + N\), where \text{etr} denotes the expected value of the trace and where \(S\) is a scaled estimate of \(\Sigma\), defined in Supplementary Material B. We then use this value to sample \(\widetilde{\lambda}\) from its full conditional, \begin{equation}
\widetilde{\lambda} \mid \Sigma, \widetilde{\alpha^\ast} \sim N_{D, K}(\widehat{L_2}, (X^\prime X + I_D)^{-1} \otimes \Sigma)
\end{equation}

\noindent a matrix variate normal distribution (see \citet{gupta1999matrix} or Supplementary Material C), where \(\widehat{L_2} = (X^\prime X + I_D)^{-1} X^\prime \widetilde{\alpha^\ast}\) is an estimate of \(\widetilde{\lambda}\), and \(X = (X_1^\prime, \ldots, X_n^\prime)^\prime \in \mathbb{R}^{N \times D}\). We note that the choice of the hyperparameter \(v_0 = K + 1\) led to the above inverse Wishart, which ``has the effect of setting a uniform distribution on the individual [covariance] parameters (that is, they are assumed equally likely to take on any value between -1 and 1)'' \citep[page 286]{gelman2007data}.

See Supplementary Material B for details on this step.

\subsection{Sparsity matrix related parameter}

We sample \(\omega\) from its full conditional: \begin{equation}
  \omega \mid \delta \sim \text{Beta}\left(\sum_{j \in [J], h \in [H]} \delta_{hj} + \omega_0,\, HJ - \sum_{j \in [J], h \in [H]} \delta_{hj} + \omega_1\right).
\end{equation}

\subsection{Initializations}

We initialize \(\alpha\) as follows. First, we perform a non-negative matrix factorization \citep{paatero1994positive, lee2000algorithms} on the response data \(Y\) specifying the number of components as \(K\), which yields a matrix of ``projected data'' that is \(N \times K\). We then for each dimension (column) of the projected data perform a \(K\)-means clustering \citep{murphy2012machine}, where the number of means for each dimension is \(L\), the number of levels of each latent attribute.

We find initial values of \(\beta\) and \(\lambda\) by finding close solutions to the systems \(d \beta = Y\) and \(x \lambda = \alpha\) respectively. This is accomplished using the \texttt{solve} function of the Armadillo C++ library (\cite{sanderson2016armadillo,sanderson2019practical}, which chooses an appropriate matrix decomposition \citep{sanderson2020adaptive}. In the case of \(\beta\), following this step, we set any negative elements to zero that appear in the above close solution so that the monotonicity condition \eqref{monotoncond} is satisfied for this chosen starting value.

\subsection{Software}

Our algorithm is implemented mostly using the Armadillo C++ library. We also make use of the scikit-learn Python package for non-negative matrix factorization and \(K\)-means clustering, and code from the PyMC package for the Geweke convergence test \cite{geweke1992evaluating}. A Code Availability statement is provided later in this manuscript. For information regarding the runtime of the algorithm, please see the Application section.

\section{Simulation studies}

\begin{table}
\small
\begin{center}
\caption{\label{tab:sim1results} Results of first simulation study}
\vspace{0.5\baselineskip}
\begin{threeparttable}[t]
\begin{tabular}{r|rlrr|lrrrr}
\toprule
\(N\) & \(J\) & \(K\) & \(L\) & \(\rho\) & \(\gamma\) & \(\eta\) & \(R\) & \(\lambda\) & \(\alpha_n\) \\
\midrule
500 & 15 & 2 & 2 & 0.000 &  & 0.014 & 0.031 & 0.079 & 0.995 \\
500 & 15 & 2 & 3 & 0.000 & 0.068 & 0.030 & 0.062 & 0.126 & 0.936 \\
500 & 25 & 3 & 2 & 0.000 &  & 0.014 & 0.043 & 0.074 & 0.993 \\
500 & 25 & 3 & 3 & 0.000 & 0.066 & 0.020 & 0.037 & 0.083 & 0.948 \\
500 & 45 & 4 & 2 & 0.000 &  & 0.016 & 0.052 & 0.082 & 0.986 \\
500 & 15 & 2 & 2 & 0.250 &  & 0.014 & 0.030 & 0.075 & 0.996 \\
500 & 15 & 2 & 3 & 0.250 & 0.061 & 0.017 & 0.023 & 0.071 & 0.992 \\
500 & 25 & 3 & 2 & 0.250 &  & 0.015 & 0.041 & 0.081 & 0.992 \\
500 & 25 & 3 & 3 & 0.250 & 0.105 & 0.041 & 0.082 & 0.106 & 0.814 \\
500 & 45 & 4 & 2 & 0.250 &  & 0.015 & 0.042 & 0.078 & 0.998 \\
500 & 15 & 2 & 2 & 0.500 &  & 0.015 & 0.026 & 0.076 & 0.996 \\
500 & 15 & 2 & 3 & 0.500 & 0.060 & 0.018 & 0.022 & 0.071 & 0.991 \\
500 & 25 & 3 & 2 & 0.500 &  & 0.042 & 0.178 & 0.175 & 0.878 \\
500 & 25 & 3 & 3 & 0.500 & 0.433 & 0.078 & 0.281 & 0.229 & 0.365 \\
500 & 45 & 4 & 2 & 0.500 &  & 0.031 & 0.120 & 0.138 & 0.872 \\
1500 & 15 & 2 & 2 & 0.000 &  & 0.008 & 0.019 & 0.051 & 0.996 \\
1500 & 15 & 2 & 3 & 0.000 & 0.042 & 0.019 & 0.039 & 0.087 & 0.953 \\
1500 & 25 & 3 & 2 & 0.000 &  & 0.008 & 0.025 & 0.052 & 0.996 \\
1500 & 25 & 3 & 3 & 0.000 & 0.039 & 0.012 & 0.021 & 0.057 & 0.967 \\
1500 & 45 & 4 & 2 & 0.000 &  & 0.012 & 0.035 & 0.063 & 0.975 \\
1500 & 15 & 2 & 2 & 0.250 &  & 0.008 & 0.016 & 0.052 & 0.996 \\
1500 & 15 & 2 & 3 & 0.250 & 0.034 & 0.009 & 0.014 & 0.047 & 0.993 \\
1500 & 25 & 3 & 2 & 0.250 &  & 0.008 & 0.021 & 0.053 & 0.996 \\
1500 & 25 & 3 & 3 & 0.250 & 0.082 & 0.032 & 0.070 & 0.087 & 0.834 \\
1500 & 45 & 4 & 2 & 0.250 &  & 0.008 & 0.026 & 0.054 & 0.998 \\
1500 & 15 & 2 & 2 & 0.500 &  & 0.008 & 0.014 & 0.053 & 0.997 \\
1500 & 15 & 2 & 3 & 0.500 & 0.031 & 0.010 & 0.013 & 0.052 & 0.992 \\
1500 & 25 & 3 & 2 & 0.500 &  & 0.033 & 0.146 & 0.138 & 0.897 \\
1500 & 25 & 3 & 3 & 0.500 & 0.462 & 0.078 & 0.295 & 0.243 & 0.328 \\
1500 & 45 & 4 & 2 & 0.500 &  & 0.025 & 0.093 & 0.119 & 0.859 \\
3000 & 15 & 2 & 2 & 0.000 &  & 0.006 & 0.013 & 0.040 & 0.996 \\
3000 & 15 & 2 & 3 & 0.000 & 0.047 & 0.025 & 0.064 & 0.115 & 0.912 \\
3000 & 25 & 3 & 2 & 0.000 &  & 0.006 & 0.017 & 0.045 & 0.997 \\
3000 & 25 & 3 & 3 & 0.000 & 0.025 & 0.008 & 0.015 & 0.042 & 0.984 \\
3000 & 45 & 4 & 2 & 0.000 &  & 0.009 & 0.027 & 0.053 & 0.977 \\
3000 & 15 & 2 & 2 & 0.250 &  & 0.006 & 0.013 & 0.041 & 0.996 \\
3000 & 15 & 2 & 3 & 0.250 & 0.026 & 0.007 & 0.011 & 0.038 & 0.993 \\
3000 & 25 & 3 & 2 & 0.250 &  & 0.006 & 0.017 & 0.046 & 0.993 \\
3000 & 25 & 3 & 3 & 0.250 & 0.091 & 0.032 & 0.066 & 0.084 & 0.797 \\
3000 & 45 & 4 & 2 & 0.250 &  & 0.006 & 0.021 & 0.047 & 0.995 \\
3000 & 15 & 2 & 2 & 0.500 &  & 0.006 & 0.010 & 0.043 & 0.997 \\
3000 & 15 & 2 & 3 & 0.500 & 0.022 & 0.007 & 0.008 & 0.040 & 0.992 \\
3000 & 25 & 3 & 2 & 0.500 &  & 0.038 & 0.179 & 0.166 & 0.864 \\
3000 & 25 & 3 & 3 & 0.500 & 0.400 & 0.062 & 0.296 & 0.202 & 0.390 \\
3000 & 45 & 4 & 2 & 0.500 &  & 0.016 & 0.060 & 0.081 & 0.911 \\
\bottomrule
\end{tabular}
  \begin{tablenotes}
    \footnotesize
    \item Values displayed for all columns except \(\alpha_n\) are the average, taken over all elements of the parameter, of the mean absolute error of estimation of that element over all replications. The values displayed for \(\alpha_n\) are the percentage of draws for which the vector \(\alpha_n\) was drawn correctly.
  \end{tablenotes}
  \end{threeparttable}
\end{center}
\end{table}

\begin{table}
\renewcommand\thetable{2}
\small
\begin{center}
\caption{Results of first simulation study (contd.)}
\vspace{0.5\baselineskip}
\begin{threeparttable}[t]
\begin{tabular}{r|rlrr|rrrrrr}
\toprule
\(N\) & \(J\) & \(K\) & \(L\) & \(\rho\) & \(\beta\) & \(\delta\) & \(\delta^0\) & \(\delta^1\) & \(\beta^0\) & \(\beta^1\) \\
\midrule
500 & 15 & 2 & 2 & 0.000 & 0.093 & 0.974 & 0.921 & 1.000 & 0.037 & 0.121 \\
500 & 15 & 2 & 3 & 0.000 & 0.190 & 0.962 & 0.932 & 0.986 & 0.076 & 0.282 \\
500 & 25 & 3 & 2 & 0.000 & 0.069 & 0.987 & 0.979 & 1.000 & 0.024 & 0.136 \\
500 & 25 & 3 & 3 & 0.000 & 0.103 & 0.970 & 0.985 & 0.932 & 0.030 & 0.285 \\
500 & 45 & 4 & 2 & 0.000 & 0.060 & 0.988 & 0.985 & 0.995 & 0.021 & 0.160 \\
500 & 15 & 2 & 2 & 0.250 & 0.095 & 0.969 & 0.908 & 1.000 & 0.041 & 0.122 \\
500 & 15 & 2 & 3 & 0.250 & 0.155 & 0.989 & 0.983 & 0.994 & 0.024 & 0.259 \\
500 & 25 & 3 & 2 & 0.250 & 0.067 & 0.988 & 0.980 & 0.999 & 0.025 & 0.131 \\
500 & 25 & 3 & 3 & 0.250 & 0.158 & 0.945 & 0.956 & 0.917 & 0.070 & 0.380 \\
500 & 45 & 4 & 2 & 0.250 & 0.050 & 0.993 & 0.991 & 0.998 & 0.015 & 0.137 \\
500 & 15 & 2 & 2 & 0.500 & 0.099 & 0.969 & 0.908 & 1.000 & 0.042 & 0.128 \\
500 & 15 & 2 & 3 & 0.500 & 0.143 & 0.980 & 0.962 & 0.994 & 0.038 & 0.227 \\
500 & 25 & 3 & 2 & 0.500 & 0.198 & 0.906 & 0.856 & 0.982 & 0.172 & 0.237 \\
500 & 25 & 3 & 3 & 0.500 & 0.282 & 0.873 & 0.862 & 0.900 & 0.189 & 0.515 \\
500 & 45 & 4 & 2 & 0.500 & 0.117 & 0.952 & 0.943 & 0.973 & 0.071 & 0.234 \\
1500 & 15 & 2 & 2 & 0.000 & 0.053 & 0.988 & 0.964 & 1.000 & 0.015 & 0.072 \\
1500 & 15 & 2 & 3 & 0.000 & 0.139 & 0.977 & 0.967 & 0.986 & 0.040 & 0.218 \\
1500 & 25 & 3 & 2 & 0.000 & 0.036 & 0.994 & 0.990 & 1.000 & 0.009 & 0.076 \\
1500 & 25 & 3 & 3 & 0.000 & 0.069 & 0.980 & 0.990 & 0.956 & 0.016 & 0.202 \\
1500 & 45 & 4 & 2 & 0.000 & 0.041 & 0.987 & 0.983 & 0.997 & 0.018 & 0.097 \\
1500 & 15 & 2 & 2 & 0.250 & 0.051 & 0.990 & 0.969 & 1.000 & 0.016 & 0.069 \\
1500 & 15 & 2 & 3 & 0.250 & 0.115 & 0.991 & 0.992 & 0.990 & 0.010 & 0.199 \\
1500 & 25 & 3 & 2 & 0.250 & 0.033 & 0.995 & 0.992 & 1.000 & 0.008 & 0.071 \\
1500 & 25 & 3 & 3 & 0.250 & 0.120 & 0.955 & 0.963 & 0.936 & 0.052 & 0.293 \\
1500 & 45 & 4 & 2 & 0.250 & 0.026 & 0.997 & 0.995 & 1.000 & 0.007 & 0.076 \\
1500 & 15 & 2 & 2 & 0.500 & 0.053 & 0.991 & 0.972 & 1.000 & 0.017 & 0.070 \\
1500 & 15 & 2 & 3 & 0.500 & 0.101 & 0.987 & 0.982 & 0.991 & 0.016 & 0.168 \\
1500 & 25 & 3 & 2 & 0.500 & 0.143 & 0.931 & 0.899 & 0.979 & 0.120 & 0.176 \\
1500 & 25 & 3 & 3 & 0.500 & 0.257 & 0.866 & 0.848 & 0.911 & 0.188 & 0.433 \\
1500 & 45 & 4 & 2 & 0.500 & 0.089 & 0.956 & 0.945 & 0.982 & 0.057 & 0.171 \\
3000 & 15 & 2 & 2 & 0.000 & 0.035 & 0.996 & 0.987 & 1.000 & 0.007 & 0.049 \\
3000 & 15 & 2 & 3 & 0.000 & 0.156 & 0.971 & 0.963 & 0.978 & 0.056 & 0.235 \\
3000 & 25 & 3 & 2 & 0.000 & 0.023 & 0.997 & 0.995 & 1.000 & 0.004 & 0.052 \\
3000 & 25 & 3 & 3 & 0.000 & 0.052 & 0.985 & 0.993 & 0.965 & 0.009 & 0.159 \\
3000 & 45 & 4 & 2 & 0.000 & 0.033 & 0.988 & 0.985 & 0.996 & 0.015 & 0.077 \\
3000 & 15 & 2 & 2 & 0.250 & 0.035 & 0.993 & 0.979 & 1.000 & 0.009 & 0.047 \\
3000 & 15 & 2 & 3 & 0.250 & 0.102 & 0.990 & 0.993 & 0.988 & 0.006 & 0.179 \\
3000 & 25 & 3 & 2 & 0.250 & 0.025 & 0.996 & 0.993 & 1.000 & 0.007 & 0.053 \\
3000 & 25 & 3 & 3 & 0.250 & 0.112 & 0.955 & 0.961 & 0.939 & 0.048 & 0.276 \\
3000 & 45 & 4 & 2 & 0.250 & 0.020 & 0.995 & 0.994 & 0.998 & 0.006 & 0.058 \\
3000 & 15 & 2 & 2 & 0.500 & 0.037 & 0.992 & 0.977 & 1.000 & 0.010 & 0.051 \\
3000 & 15 & 2 & 3 & 0.500 & 0.086 & 0.988 & 0.988 & 0.988 & 0.009 & 0.148 \\
3000 & 25 & 3 & 2 & 0.500 & 0.167 & 0.905 & 0.861 & 0.972 & 0.156 & 0.183 \\
3000 & 25 & 3 & 3 & 0.500 & 0.199 & 0.895 & 0.877 & 0.940 & 0.147 & 0.328 \\
3000 & 45 & 4 & 2 & 0.500 & 0.057 & 0.971 & 0.965 & 0.988 & 0.033 & 0.119 \\
\bottomrule
\end{tabular}
  \begin{tablenotes}
    \footnotesize
    \item Values displayed for \(\beta\) parameters are the average, taken over all elements of the parameter, of the mean absolute error of estimation of each element over all replications. Values displayed for \(\delta\) parameters are the average, taken over all elements of the parameter, of the recovery accuracy of each element over all replications.
  \end{tablenotes}
  \end{threeparttable}
\end{center}
\end{table}

\begin{table}
\renewcommand\thetable{3}
\small
\begin{center}
\caption{\label{tab:sim2results} Results of second simulation study}
\vspace{0.5\baselineskip}
\begin{threeparttable}[t]
\begin{tabular}{rrrlr|lrrrr}
\toprule
\(N\) & \(J\) & \(K\) & \(L\) & \(\rho\) & \(\gamma\) & \(\eta\) & \(R\) & \(\lambda\) & \(\alpha_n\) \\
\midrule
3000 & 45 & 4 & 3 & 0.000 & 0.044 & 0.015 & 0.030 & 0.053 & 0.917\\
3000 & 70 & 5 & 2 & 0.000 &  & 0.009 & 0.026 & 0.051 & 0.976 \\
3000 & 70 & 5 & 3 & 0.000 & 0.027 & 0.013 & 0.024 & 0.047 & 0.957 \\
3000 & 45 & 4 & 3 & 0.250 & 0.058 & 0.024 & 0.053 & 0.080 & 0.846 \\
3000 & 70 & 5 & 2 & 0.250 &  & 0.007 & 0.022 & 0.046 & 0.985 \\
3000 & 70 & 5 & 3 & 0.250 & 0.056 & 0.024 & 0.053 & 0.064 & 0.843 \\
3000 & 45 & 4 & 3 & 0.500 & 0.115 & 0.061 & 0.163 & 0.127 & 0.655 \\
3000 & 70 & 5 & 2 & 0.500 &  & 0.017 & 0.067 & 0.076 & 0.914 \\
3000 & 70 & 5 & 3 & 0.500 & 0.081 & 0.035 & 0.100 & 0.084 & 0.791 \\
5000 & 45 & 4 & 3 & 0.000 & 0.040 & 0.014 & 0.026 & 0.050 & 0.907 \\
5000 & 70 & 5 & 2 & 0.000 &  & 0.008 & 0.022 & 0.049 & 0.972 \\
5000 & 70 & 5 & 3 & 0.000 & 0.025 & 0.011 & 0.020 & 0.044 & 0.946 \\
5000 & 45 & 4 & 3 & 0.250 & 0.050 & 0.022 & 0.049 & 0.073 & 0.857 \\
5000 & 70 & 5 & 2 & 0.250 &  & 0.005 & 0.015 & 0.041 & 0.993 \\
5000 & 70 & 5 & 3 & 0.250 & 0.062 & 0.023 & 0.053 & 0.059 & 0.819 \\
5000 & 45 & 4 & 3 & 0.500 & 0.121 & 0.064 & 0.167 & 0.126 & 0.620 \\
5000 & 70 & 5 & 2 & 0.500 &  & 0.018 & 0.063 & 0.088 & 0.906 \\
5000 & 70 & 5 & 3 & 0.500 & 0.084 & 0.036 & 0.099 & 0.087 & 0.756 \\
\bottomrule
\end{tabular}
  \begin{tablenotes}
    \footnotesize
    \item Values displayed for all columns except \(\alpha_n\) are the average, taken over all elements of the parameter, of the mean absolute error of estimation of each element over all replications. The values displayed for \(\alpha_n\) are the percentage of draws for which the vector \(\alpha_n\) was drawn correctly.
  \end{tablenotes}
  \end{threeparttable}
\end{center}
\end{table}

\begin{table}
\renewcommand\thetable{3}
\small
\begin{center}
\caption{\label{simtwopartone} Results of second simulation study (contd.)}
\vspace{0.5\baselineskip}
\begin{threeparttable}[t]
\begin{tabular}{rrrlr|rrrrrr}
\toprule
\(N\) & \(J\) & \(K\) & \(L\) & \(\rho\) & \(\beta\) & \(\delta\) & \(\delta^0\) & \(\delta^1\) & \(\beta^0\) & \(\beta^1\) \\
\midrule
3000 & 45 & 4 & 3 & 0.000 & 0.046 & 0.984 & 0.987 & 0.972 & 0.014 & 0.182 \\
3000 & 70 & 5 & 2 & 0.000 & 0.021 & 0.994 & 0.993 & 0.997 & 0.007 & 0.077 \\
3000 & 70 & 5 & 3 & 0.000 & 0.033 & 0.990 & 0.994 & 0.965 & 0.007 & 0.199 \\
3000 & 45 & 4 & 3 & 0.250 & 0.060 & 0.974 & 0.975 & 0.967 & 0.027 & 0.199 \\
3000 & 70 & 5 & 2 & 0.250 & 0.017 & 0.997 & 0.996 & 0.999 & 0.004 & 0.067 \\
3000 & 70 & 5 & 3 & 0.250 & 0.047 & 0.979 & 0.984 & 0.952 & 0.018 & 0.230 \\
3000 & 45 & 4 & 3 & 0.500 & 0.136 & 0.921 & 0.922 & 0.919 & 0.091 & 0.327 \\
3000 & 70 & 5 & 2 & 0.500 & 0.040 & 0.978 & 0.975 & 0.991 & 0.025 & 0.098 \\
3000 & 70 & 5 & 3 & 0.500 & 0.085 & 0.966 & 0.972 & 0.928 & 0.047 & 0.328 \\
5000 & 45 & 4 & 3 & 0.000 & 0.040 & 0.986 & 0.987 & 0.980 & 0.013 & 0.153 \\
5000 & 70 & 5 & 2 & 0.000 & 0.020 & 0.993 & 0.991 & 0.998 & 0.008 & 0.066 \\
5000 & 70 & 5 & 3 & 0.000 & 0.025 & 0.993 & 0.995 & 0.979 & 0.005 & 0.155 \\
5000 & 45 & 4 & 3 & 0.250 & 0.052 & 0.976 & 0.977 & 0.975 & 0.024 & 0.168 \\
5000 & 70 & 5 & 2 & 0.250 & 0.012 & 0.997 & 0.996 & 1.000 & 0.003 & 0.049 \\
5000 & 70 & 5 & 3 & 0.250 & 0.042 & 0.979 & 0.981 & 0.968 & 0.019 & 0.191 \\
5000 & 45 & 4 & 3 & 0.500 & 0.137 & 0.916 & 0.914 & 0.926 & 0.093 & 0.319 \\
5000 & 70 & 5 & 2 & 0.500 & 0.041 & 0.972 & 0.968 & 0.988 & 0.029 & 0.091 \\
5000 & 70 & 5 & 3 & 0.500 & 0.075 & 0.964 & 0.967 & 0.946 & 0.041 & 0.297 \\
\bottomrule
\end{tabular}
  \begin{tablenotes}
    \footnotesize
    \item Values displayed for \(\beta\) parameters are the average, taken over all elements of the parameter, of the mean absolute error of estimation of each element over all replications. Values displayed for \(\delta\) parameters are the average, taken over all elements of the parameter, of the recovery accuracy of each element over all replications.
  \end{tablenotes}
  \end{threeparttable}
\end{center}
\end{table}

\begin{table}
\renewcommand\thetable{4}
\small
\begin{center}
\caption{\label{sim1mcse} Average Monte Carlo errors for first simulation study}
\begin{tabular}{r|rlrr|rrrlrr}
\toprule
\(N\) & \(J\) & \(K\) & \(L\) & \(\rho\) & \(\eta\) & \(\beta\) & \(\delta\) & \(\gamma\) & \(R\) & \(\lambda\) \\
\midrule
500 & 15 & 2 & 2 & 0.000 & 0.001 & 0.008 & 0.009 &  & 0.005 & 0.002 \\
500 & 15 & 2 & 3 & 0.000 & 0.005 & 0.019 & 0.011 & 0.005 & 0.012 & 0.007 \\
500 & 25 & 3 & 2 & 0.000 & 0.001 & 0.007 & 0.007 &  & 0.005 & 0.003 \\
500 & 25 & 3 & 3 & 0.000 & 0.003 & 0.013 & 0.011 & 0.010 & 0.008 & 0.003 \\
500 & 45 & 4 & 2 & 0.000 & 0.002 & 0.009 & 0.008 &  & 0.006 & 0.005 \\
500 & 15 & 2 & 2 & 0.250 & 0.001 & 0.008 & 0.009 &  & 0.005 & 0.002 \\
500 & 15 & 2 & 3 & 0.250 & 0.001 & 0.012 & 0.006 & 0.005 & 0.005 & 0.002 \\
500 & 25 & 3 & 2 & 0.250 & 0.001 & 0.008 & 0.008 &  & 0.005 & 0.004 \\
500 & 25 & 3 & 3 & 0.250 & 0.008 & 0.022 & 0.017 & 0.013 & 0.010 & 0.010 \\
500 & 45 & 4 & 2 & 0.250 & 0.002 & 0.005 & 0.005 &  & 0.005 & 0.003 \\
500 & 15 & 2 & 2 & 0.500 & 0.001 & 0.008 & 0.009 &  & 0.005 & 0.002 \\
500 & 15 & 2 & 3 & 0.500 & 0.002 & 0.011 & 0.009 & 0.004 & 0.004 & 0.001 \\
500 & 25 & 3 & 2 & 0.500 & 0.005 & 0.028 & 0.021 &  & 0.020 & 0.023 \\
500 & 25 & 3 & 3 & 0.500 & 0.008 & 0.035 & 0.029 & 0.040 & 0.022 & 0.022 \\
500 & 45 & 4 & 2 & 0.500 & 0.005 & 0.022 & 0.017 &  & 0.020 & 0.018 \\
1500 & 15 & 2 & 2 & 0.000 & 0.001 & 0.004 & 0.006 &  & 0.003 & 0.001 \\
1500 & 15 & 2 & 3 & 0.000 & 0.004 & 0.014 & 0.009 & 0.004 & 0.010 & 0.006 \\
1500 & 25 & 3 & 2 & 0.000 & 0.001 & 0.003 & 0.005 &  & 0.002 & 0.002 \\
1500 & 25 & 3 & 3 & 0.000 & 0.004 & 0.011 & 0.009 & 0.007 & 0.006 & 0.003 \\
1500 & 45 & 4 & 2 & 0.000 & 0.003 & 0.010 & 0.008 &  & 0.007 & 0.005 \\
1500 & 15 & 2 & 2 & 0.250 & 0.001 & 0.004 & 0.006 &  & 0.002 & 0.001 \\
1500 & 15 & 2 & 3 & 0.250 & 0.001 & 0.008 & 0.005 & 0.002 & 0.002 & 0.001 \\
1500 & 25 & 3 & 2 & 0.250 & 0.001 & 0.003 & 0.004 &  & 0.002 & 0.002 \\
1500 & 25 & 3 & 3 & 0.250 & 0.008 & 0.019 & 0.014 & 0.011 & 0.011 & 0.010 \\
1500 & 45 & 4 & 2 & 0.250 & 0.001 & 0.003 & 0.003 &  & 0.003 & 0.002 \\
1500 & 15 & 2 & 2 & 0.500 & 0.001 & 0.004 & 0.005 &  & 0.003 & 0.001 \\
1500 & 15 & 2 & 3 & 0.500 & 0.001 & 0.008 & 0.007 & 0.002 & 0.002 & 0.001 \\
1500 & 25 & 3 & 2 & 0.500 & 0.004 & 0.021 & 0.016 &  & 0.017 & 0.021 \\
1500 & 25 & 3 & 3 & 0.500 & 0.009 & 0.036 & 0.028 & 0.043 & 0.024 & 0.023 \\
1500 & 45 & 4 & 2 & 0.500 & 0.006 & 0.021 & 0.017 &  & 0.021 & 0.017 \\
3000 & 15 & 2 & 2 & 0.000 & 0.001 & 0.003 & 0.003 &  & 0.002 & 0.001 \\
3000 & 15 & 2 & 3 & 0.000 & 0.005 & 0.014 & 0.008 & 0.004 & 0.012 & 0.008 \\
3000 & 25 & 3 & 2 & 0.000 & 0.000 & 0.002 & 0.003 &  & 0.002 & 0.001 \\
3000 & 25 & 3 & 3 & 0.000 & 0.002 & 0.008 & 0.007 & 0.003 & 0.002 & 0.002 \\
3000 & 45 & 4 & 2 & 0.000 & 0.003 & 0.011 & 0.009 &  & 0.007 & 0.004 \\
3000 & 15 & 2 & 2 & 0.250 & 0.001 & 0.003 & 0.004 &  & 0.002 & 0.001 \\
3000 & 15 & 2 & 3 & 0.250 & 0.001 & 0.007 & 0.005 & 0.002 & 0.002 & 0.001 \\
3000 & 25 & 3 & 2 & 0.250 & 0.000 & 0.004 & 0.004 &  & 0.003 & 0.002 \\
3000 & 25 & 3 & 3 & 0.250 & 0.007 & 0.017 & 0.012 & 0.011 & 0.009 & 0.009 \\
3000 & 45 & 4 & 2 & 0.250 & 0.001 & 0.004 & 0.004 &  & 0.003 & 0.003 \\
3000 & 15 & 2 & 2 & 0.500 & 0.001 & 0.003 & 0.004 &  & 0.002 & 0.001 \\
3000 & 15 & 2 & 3 & 0.500 & 0.001 & 0.006 & 0.006 & 0.002 & 0.002 & 0.001 \\
3000 & 25 & 3 & 2 & 0.500 & 0.005 & 0.025 & 0.019 &  & 0.020 & 0.022 \\
3000 & 25 & 3 & 3 & 0.500 & 0.007 & 0.032 & 0.024 & 0.034 & 0.023 & 0.021 \\
3000 & 45 & 4 & 2 & 0.500 & 0.004 & 0.016 & 0.013 &  & 0.014 & 0.013 \\
\bottomrule
\end{tabular}
\end{center}
\end{table}

\begin{table}
\renewcommand\thetable{5}
\small
\begin{center}
\caption{\label{tab:sim2mcse} Average Monte Carlo errors for second simulation study}
\begin{tabular}{r|rlrr|rrrlrr}
\toprule
\(N\) &  \(J\) & \(K\) & \(L\) &  \(\rho\) & \(\eta\) & \(\beta\) & \(\delta\) & \(\gamma\) & \(R\) & \(\lambda\) \\
\midrule
3000 & 45 & 4 & 3 & 0.000 & 0.004 & 0.010 & 0.008 & 0.009 & 0.005 & 0.006 \\
3000 & 70 & 5 & 2 & 0.000 & 0.003 & 0.005 & 0.005 &       & 0.006 & 0.004 \\
3000 & 70 & 5 & 3 & 0.000 & 0.004 & 0.006 & 0.005 & 0.003 & 0.003 & 0.004 \\
3000 & 45 & 4 & 3 & 0.250 & 0.006 & 0.013 & 0.011 & 0.008 & 0.010 & 0.008 \\
3000 & 70 & 5 & 2 & 0.250 & 0.002 & 0.004 & 0.003 &       & 0.004 & 0.003 \\
3000 & 70 & 5 & 3 & 0.250 & 0.007 & 0.010 & 0.009 & 0.009 & 0.008 & 0.009 \\
3000 & 45 & 4 & 3 & 0.500 & 0.010 & 0.027 & 0.022 & 0.014 & 0.016 & 0.017 \\
3000 & 70 & 5 & 2 & 0.500 & 0.004 & 0.013 & 0.011 &       & 0.014 & 0.015 \\
3000 & 70 & 5 & 3 & 0.500 & 0.008 & 0.034 & 0.013 & 0.013 & 0.009 & 0.014 \\
5000 & 45 & 4 & 3 & 0.000 & 0.004 & 0.009 & 0.008 & 0.008 & 0.005 & 0.006 \\
5000 & 70 & 5 & 2 & 0.000 & 0.003 & 0.007 & 0.006 &  & 0.008 & 0.004 \\
5000 & 70 & 5 & 3 & 0.000 & 0.004 & 0.004 & 0.003 & 0.004 & 0.003 & 0.004 \\
5000 & 45 & 4 & 3 & 0.250 & 0.006 & 0.012 & 0.010 & 0.008 & 0.009 & 0.009 \\
5000 & 70 & 5 & 2 & 0.250 & 0.001 & 0.002 & 0.003 & & 0.003 & 0.001 \\
5000 & 70 & 5 & 3 & 0.250 & 0.007 & 0.011 & 0.009 & 0.011 & 0.006 & 0.009 \\
5000 & 45 & 4 & 3 & 0.500 & 0.010 & 0.028 & 0.022 & 0.014 & 0.014 & 0.016 \\
5000 & 70 & 5 & 2 & 0.500 & 0.005 & 0.015 & 0.014 & & 0.020 & 0.017 \\
5000 & 70 & 5 & 3 & 0.500 & 0.008 & 0.029 & 0.013 & 0.012 & 0.010 & 0.013 \\
\bottomrule
\end{tabular}
\end{center}
\end{table}

In our first simulation study, we study 45 scenarios, which differ in the combination of \(N, J, K\) and \(L\) (as well as in a value \(\rho\) which is used for the off-diagonal values of parameter \(R\)) as shown in Table \ref{tab:sim1results}. For each combination of \(K\) and \(L\) we choose \(J\), \(\beta\), \(\kappa\), and \(\gamma\) such that for each item, the pattern of class-conditional response probabilities differs across classes, making it possible to distinguish amongst classes (the parameter values are available in a repository whose link is included in the Code Availability statement towards the end of this manuscript). We also generate data so that there are at least some respondents in each class. For all scenarios, we set hyperparameters to the following values: \(\sigma_{\beta}^2 = 2.0\), \(\omega_0 = 0.5\) and \(\omega_1 = 0.5\), \(a = 1000^{-1}\), and \(v_0 = K + 1\). We tuned \(\sigma_{\kappa}^2\) for each sample size so that its acceptance rate was approximately 40\%. The order of the model was set to 2 for all scenarios.

The generated data consists both of covariates \(X\) and response data \(Y\). \(Y\) is generated from the model and data-generating parameters, and \(X\) is generated to mimic the covariates in the application to follow, where one is age and the other is sex (we also include a column of ones to fit an intercept). Age data is generated from a categorical random variable with pre-specified probabilities for three age ranges, then by sampling from a truncated normal on the appropriate range and converting each sampled value to an integer. Data for the sex of the respondents is generated by drawing from a Bernoulli with parameter 0.6 (we code male as 0, female as 1).

We perform 100 replications for each scenario: namely for the set of data-generating quantities (e.g. \(J\)) and parameters we generate 100 datasets and run our algorithm on each of the 100, and measure parameter recovery in each case. The data on parameter recovery for each replication is aggregated for reporting purposes in Table \ref{tab:sim1results}. We used a chain length after burn-in of 10,000 iterations, and a burn-in period of 6,000 iterations, which resulted in fewer than 5\% of the elements of any of our matrix parameters failing the Geweke test for convergence. We evaluate recovery of \(\gamma, \eta, R, \lambda, \beta\) and \(\delta\), where \(\eta\) refers to the set of class-conditional item response probabilities.

For each scenario, recovery for each element of \(\gamma, \eta, R, \lambda\), and \(\beta\) is evaluated by computing the absolute error of estimation for each replication and taking the mean value of these across all replications to produce the mean absolute error of estimation for each element. For any one element of the above parameters (referring to that element as \(\theta\) in this discussion), the absolute error of estimation of \(\theta\) for replication \(r\) is \(\text{AE}_{r}(\theta, \widehat{\theta}^{(r)}) = |\theta - \widehat{\theta}^{(r)}|\) where \(\widehat{\theta}^{(r)}\) is the estimate of \(\theta\) for replication \(r\), and the mean absolute error (MAE) across all 100 replications for \(\theta\) is the average of these. For these parameters we use the estimate of the posterior mean \(\widehat{\theta}^{(r)} = (1 / S) \sum_{s=1}^S \theta^{(s, r)}\), where \(\theta^{(s, r)}\) is the \(s\)th draw of \(\theta\) from the Markov chain for replication \(r\) in the post-burn-in phase and \(S\) is the number of post-burn-in draws. For elements of \(\delta\) we use correctness of estimation rather than absolute error: the recovery accuracy for any element \(\theta\) is the average of \(I(\theta = \widehat{\theta}^{(r)})\) over all replications. We use \(\widehat{\theta}^{(r)} = I\left((1/S) \sum_{s=1}^S \theta^{(r, s)} > 0.5\right)\), the estimate of the posterior median. We report the average of these recovery accuracy values across all elements of \(\delta\) in the Table.

In Table \ref{tab:sim1results}, for each of \(\gamma, \eta, R, \lambda\) and \(\beta\) we report the average mean absolute error of estimation taken over all its elements, and for \(\delta\) we report the average recovery accuracy. In addition to the above, we compare recovery for active and inactive coefficients of \(\beta\) separately and report those in Table \ref{tab:sim1results} as well. Average recovery accuracy across all \(\delta_{hj}\) for which \(\delta_{hj} = 0\) is reported under the header \(\delta^0\) and average MAE of the corresponding elements of \(\beta\) (the inactive coefficients) is reported under \(\beta^0\). Similarly, average recovery accuracy across all \(\delta_{hj}\) for which \(\delta_{hj} = 1\) is reported under \(\delta^1\), and average mean absolute error of the corresponding elements of \(\beta\) (the active coefficients) is reported under \(\beta^1\).

Regarding the results shown in Table \ref{tab:sim1results}, for each combination of \(J, K, L\) and \(\rho\) except one, we observe that the average MAE of each parameter decreases as \(N\) increases. For one combination, namely \(K = 3\) with \(L = 3\) and \(\rho = 0.5\), we do not observe this behavior, since the correlation of the attributes of the latent state is rather high, which when combined with the increased number of attributes makes parameter recovery more difficult. We observe that the higher the classication accuracy of \(\alpha_n\) for a scenario, the better is the parameter recovery.

We performed a second simulation study to examine the performance for combinations of \(K\) and \(L\) that have a higher number of possible values for the latent state, the results of which are shown in Table \ref{tab:sim2results}. We consider three additional combinations of \(K\) and \(L\): \(K = 4, L = 3\), \(K = 5, L = 2\) and \(K = 5, L = 3\). We use sample sizes of 3,000 and 5,000, where 3,000 was the largest sample size of the first simulation study (using this sample size gives an average of at least 12 people per class). For most scenarios, increasing sample size 3,000 to 5,000 results in a slight improvement in parameter recovery. It should be noted that our metric of MAE is sensitive to outliers, so that for a given set of \(J, K, L\), even if most replications perform well for both the \(N = 3,000\) and \(N = 5,000\) cases, if just one additional replication performs poorly for \(N = 5,000\) that will lead to a worse value in our performance metric even if most replications are performing better with that larger sample size.

For both simulation studies, for each parameter, we also calculate the Monte Carlo error of the estimates for each element across all replications, and report the average across elements in Tables \ref{sim1mcse} and \ref{tab:sim2mcse}. Our performance metrics have low average Monte Carlo error for most scenarios. For certain scenarios we see higher values, which reflect the presence of a small number of poorly performing replications.

\section{Application}

\subsection{Data}

We apply our model to the Hamilton Rating Scale for Depression \citep{hamilton1960rating, hamilton1967development} from a study entitled Sequenced Treatment Alternatives to Relieve Depression, or STAR*D \citep{fava2003background,rush2004sequenced}, which was sponsored by the National Institute of Mental Health (NIMH) of the United States of America. The items of the Hamilton Rating Scale for Depression, or HRSD, are described in Table \ref{tab:itemdescs}: we note that there are 17 items and that each is ordinal. Our response data was collected at the entry point of this study, for which there were 3,960 respondents with complete responses to the HRSD. The purpose of the study was to establish a more scientific basis for recommending alternative medications to patients suffering depression who did not have a positive response to the initial medication prescribed to all respondents.

\begin{table}
\renewcommand\thetable{6}
\small
\begin{center}
\caption{\label{tab:itemdescs} Item labels, descriptions, and observed frequencies}
\vspace{0.5\baselineskip}
\begin{tabular}{lllllll}
\toprule
Label & Description & 0 & 1 & 2 & 3 & 4 \\
\midrule
hsoin & Initial insomnia & 0.281 & 0.136 & 0.583 & -- & -- \\
hmnin & Middle insomnia & 0.174 & 0.288 & 0.538 & -- & -- \\
hemin & Late insomnia & 0.437 & 0.183 & 0.380 & -- & -- \\
hmdsd & Depressed mood & 0.004 & 0.053 & 0.364 & 0.464 & 0.115 \\
hpanx & Psychic anxiety & 0.035 & 0.134 & 0.474 & 0.324 & 0.033 \\
hinsg & Loss of insight & 0.906 & 0.062 & 0.032 & -- & -- \\
happt & Appetite & 0.496 & 0.273 & 0.231 & -- & -- \\
hwl & Weight loss & 0.682 & 0.206 & 0.113 & -- & -- \\
hsanx & Somatic anxiety & 0.080 & 0.301 & 0.413 & 0.181 & 0.026 \\
hhypc & Hypochondriasis & 0.411 & 0.263 & 0.232 & 0.094 & 0.000 \\
hvwsf & Guilt feelings and delusions & 0.086 & 0.161 & 0.385 & 0.359 & 0.009 \\
hsuic & Suicide & 0.371 & 0.383 & 0.218 & 0.025 & 0.003 \\
hintr & Work and interests & 0.022 & 0.092 & 0.299 & 0.525 & 0.062 \\
hengy & Somatic energy & 0.037 & 0.240 & 0.723 & -- & -- \\
hslow & Retardation (psychomotor) & 0.246 & 0.468 & 0.248 & 0.038 & -- \\
hagit & Agitation & 0.292 & 0.447 & 0.218 & 0.041 & 0.002 \\
hsex & Libido & 0.308 & 0.271 & 0.420 & -- & -- \\
\bottomrule
\end{tabular}
\end{center}
\end{table}

\subsection{Model selection}\label{sec:modelsel}

We ran a selection of models of increasing complexity, displayed in Table \ref{tab:ppcresults}, using respondents' age and sex as covariates (age was standardized, sex was coded as 0 for male and 1 for female). We restricted the choice of \(L\) to values of either \(2\) or \(3\) so that the number of classes would not grow too large too quickly as \(K\) increased. We set hyperparameters as follows: \(\sigma_{\beta}^2 = 2.0\), \(\omega_0 = 0.5\), and \(\omega_1 = 0.5\), \(a = 1000^{-1}\), and \(v_0 = K + 1\). We tuned \(\sigma_{\kappa}^2\) so that its acceptance rate was approximately 40\%. The order of the model was set to 2. For the chosen model, to check convergence we perform the check of \cite{geweke1992evaluating} on the post-burn-in section of the chain of draws of original model parameters on a per-parameter basis \cite[page 170]{gelman2011inference} and do not reject for any of these the null hypothesis that the sequence of draws has converged. We also record the time in seconds to perform 1,000 MCMC draws under each model for this dataset (the runs were performed using 8 cores of an 11th Gen Intel(R) Core(TM) i7-1160G7 processor; this processor was released in the year 2020).

We implement a posterior predictive check \citep{gelman1996posterior} in order to help guide model selection as follows. We first define a vector-valued statistic designed to capture the structure of a dataset consisting of ordinal items. As described in \citet[page 98]{bartholomew2011latent}, for each pair of items in the dataset, we count the number of times each possible combination of levels of the items in the pair occurs in the data, and build a \(p\)-dimensional vector containing these counts, where \(p\) is the total number of combinations of levels for all pairs (for the HRSD, \(p = 2298\)). We also calculate this vector for a series of draws from the posterior predictive distribution.

Similar to \cite{crespi2009bayesian}, we form two sets of distances to be compared, where the distance between vectors \(x\) and \(y\) is computed as the sum of the absolute differences between their components, and is denoted by \(d(x, y)\). The first of these is the set of distances of each of the statistics for 1,000 draws from the posterior predictive distribution to the observed data statistic, denoted \(D_{\text{obs}, \text{ppred}}\). The second of these, \(D_{\text{ppred}}\), is a set of distances between each of the 2,500 pairs of the statistics for independent draws from the posterior predictive distribution.

\begin{table}
\renewcommand\thetable{7}
\small
\begin{center}
\caption{\label{tab:ppcresults} Models, their complexity and runtime, and hypothesis test results}
\vspace{0.5\baselineskip}
\begin{threeparttable}[t]
  \begin{tabular}{ccccccc}
    \toprule
K & levels & classes & meas. params\tnote{a} & struct. params\tnote{b} & \(p\)-value\tnote{c} & benchmark\tnote{d} \\
\midrule
2 & 2 & 4 & 104 & 7 &   0.00000 & 11 \\
3 & 2 & 8 & 155 & 12 &  0.00002 & 17 \\
2 & 3 & 9 & 189 & 9 &   0.00014 & 15 \\
4 & 2 & 16 & 223 & 18 & 1.00000 & 28 \\
3 & 3 & 27 & 359 & 15 & 1.00000 & 33 \\
\bottomrule
  \end{tabular}
  \begin{tablenotes}
    \footnotesize
    \item[a] number of measurement model parameters
    \item[b] number of structural model parameters
    \item[c] \(p\)-value for Mann-Whitney U-test
    \item[d] Number of seconds to perform 1,000 MCMC draws
  \end{tablenotes}
  \end{threeparttable}
\end{center}
\end{table}

To check a particular model, inspired by \citet{crespi2009bayesian} we calculate the test statistic for the Mann-Whitney U-test \citep{mann1947test}, with the null hypothesis being that the distributions of \(D_{\text{obs}, \text{ppred}}\) and \(D_{\text{ppred}}\) are the same, and with the one-sided alternative hypothesis that \(D_{\text{obs}, \text{ppred}}\) is stochastically greater than \(D_{\text{ppred}}\) (i.e. that our statistics for the posterior predictive distribution draws are farther from the statistic for the observed data than they are from each other). We perform this test for each of the models in Table \ref{tab:ppcresults}. As seen from the \(p\)-values in the Table, the two simplest models for which we do not reject the above null hypothesis are \(K = 4, L = 2\) and \(K = 3, L = 3\).

To choose between these two models, we consider the sparsity of the estimate of the \(\beta\) matrix, where coefficients are left out if the value 0 falls into the 95\% equal-tailed credible interval. Between the two models with the highest number of classes (namely \(K = 4, L = 2\) and \(K = 3, L = 3\)), a higher proportion of the elements of \(\beta\) are removed for the \(K = 3, L = 3\) model. Summarizing the differences between these two models, although the \(K = 3, L = 3\) model has a larger number of classes and measurement model parameters, it has greater sparsity of \(\beta\), a smaller number of structural model parameters, and a lower value of the test statistic. We choose to select the \(K = 3, L = 3\) model, while acknowledging that practitioners may have domain-specific knowledge that may aid in the appropriate selection of a model in this sort of situation.

\subsection{Interpretation of fitted model}

\begin{table}
\renewcommand\thetable{8}
\small
\begin{center}
\caption{\label{tab:applicbeta1} Beta coefficients}
\vspace{0.5\baselineskip}
\begin{tabular}{lrrrrrrr}
\toprule
effect & [0 0 0] & [0 0 1] & [0 0 2] & [0 1 0] & [0 2 0] & [1 0 0] & [2 0 0] \\
\midrule
hsoin & -0.26 &  &  &  &  &  &  \\
hmnin & 0.23 &  &  &  &  &  & 1.40 \\
hemin & -0.62 &  &  &  &  &  & 1.26 \\
hmdsd & 1.98 &  &  &  &  &  &  \\
hpanx & 1.25 &  &  &  &  &  &  \\
hinsg & -1.42 &  &  &  &  &  &  \\
happt & -2.09 &  &  & 1.50 & 1.49 &  &  \\
hwl & -2.46 &  &  & 1.31 & 0.96 &  &  \\
hsanx & 0.24 &  &  &  &  & 0.92 &  \\
hhypc & -0.74 &  &  &  &  & 0.92 &  \\
hvwsf & 0.84 & 0.55 &  &  &  &  &  \\
hsuic & -0.23 &  & 0.56 &  &  &  &  \\
hintr & 1.56 &  & 0.93 &  &  &  &  \\
hengy & 1.07 &  &  &  &  &  &  \\
hslow & 0.01 &  &  &  &  &  &  \\
hagit & 0.25 &  &  &  &  &  &  \\
hsex & -0.28 &  &  &  &  &  &  \\
\bottomrule
\end{tabular}
\end{center}
\end{table}

\begin{table}
\renewcommand\thetable{8}
\small
\begin{center}
\caption{\label{tab:applicbeta2} Beta coefficients (contd.)}
\vspace{0.5\baselineskip}
\begin{tabular}{lrrrrrrr}
\toprule
effect & [0 1 1] & [0 1 2] & [0 2 1] & [0 2 2] & [1 0 1] & [1 0 2] & [1 1 0] \\
\midrule
hsoin &  &  &  &  &  &  &  \\
hmnin & 0.57 &  &  &  &  &  &  \\
hemin &  &  &  &  &  &  &  \\
hmdsd &  &  &  &  &  &  &  \\
hpanx & 0.98 &  &  &  & 0.67 &  &  \\
hinsg &  &  &  &  &  &  &  \\
happt &  &  & 1.16 &  &  &  &  \\
hwl &  &  & 0.79 &  &  &  &  \\
hsanx & 1.11 &  &  &  &  &  &  \\
hhypc & 0.49 &  &  &  &  &  &  \\
hvwsf &  &  &  &  &  &  &  \\
hsuic &  &  &  &  &  &  &  \\
hintr &  &  &  &  &  &  &  \\
hengy &  &  &  &  &  &  &  \\
hslow &  &  &  &  &  &  &  \\
hagit & 0.54 &  &  &  &  &  &  \\
hsex &  &  &  &  &  &  &  \\
\bottomrule
\end{tabular}
\end{center}
\end{table}

\begin{table}
\renewcommand\thetable{8}
\small
\begin{center}
\caption{\label{tab:applicbeta3} Beta coefficients (contd.)}
\vspace{0.5\baselineskip}
\begin{tabular}{lrrrrr}
\toprule
effect & [1 2 0] & [2 0 1] & [2 0 2] & [2 1 0] & [2 2 0] \\
\midrule
hsoin &  &  &  &  &  \\
hmnin &  &  &  &  &  \\
hemin &  &  &  &  &  \\
hmdsd & 0.86 &  &  &  &  \\
hpanx &  &  & 1.52 &  &  \\
hinsg &  &  &  &  &  \\
happt &  & 0.95 &  & 1.39 &  \\
hwl &  &  &  &  &  \\
hsanx &  &  & 1.84 &  &  \\
hhypc &  &  & 0.93 &  &  \\
hvwsf &  &  &  &  &  \\
hsuic &  &  &  &  &  \\
hintr & 1.00 & 0.58 &  &  &  \\
hengy & 0.77 &  &  &  &  \\
hslow &  & 0.42 &  &  &  \\
hagit &  &  & 0.81 &  &  \\
hsex &  &  &  &  &  \\
\bottomrule
\end{tabular}
\end{center}
\end{table}

\begin{table}
\renewcommand\thetable{9}
\small
\begin{center}
\caption{\label{tab:appliclambda} Lambda and gamma coefficients}
\vspace{0.5\baselineskip}
\begin{threeparttable}[t]
\begin{tabular}{lrrr}
\toprule
 & Attribute 1 & Attribute 2 & Attribute 3 \\
\midrule
Intercept & 0.71 & 0.49 & 0.52 \\
Female & 0.52 & -0.23 & -0.24 \\
Age & 0.41 & -0.13 & \({}^{\ast}\) -0.07 \\
\midrule
Threshold & 1.17 & 1.17 & 1.17 \\
\bottomrule
\end{tabular}
  \begin{tablenotes}
    \item[(*)] 95\% equal-tailed credible interval contains zero
  \end{tablenotes}
  \end{threeparttable}
\end{center}
\end{table}

\begin{table}
\renewcommand\thetable{10}
\small
\begin{center}
\caption{\label{tab:applicRmat} Correlation matrix coefficients}
\vspace{0.5\baselineskip}
\begin{tabular}{lrrr}
\toprule
 & Attribute 1 & Attribute 2 & Attribute 3 \\
\midrule
Attribute 1 & 1.00 & -0.23 & -0.47 \\
Attribute 2 & -0.23 & 1.00 & -0.37 \\
Attribute 3 & -0.47 & -0.37 & 1.00 \\
\bottomrule
\end{tabular}
\end{center}
\end{table}

Table \ref{tab:applicbeta1} shows the model's estimates of the elements of \(\beta\), where we have excluded elements for which the 95\% equal-tailed credible interval contains the number zero. We first examine the main effects (effects involving only a single attribute) for each level. Ignoring the ``intercept'' (all levels 0), we see that attribute 1 could be called ``anxiety'': the level 1 main effect reflects somatic anxiety and hyperchondriasis, and the level 2 main effect reflects two types of insomnia, which can happen due to anxiety. Attribute 2 could be called ``weight-related'' since the main effects for both levels reflect issues with appetite and weight loss. Attribute 3 could be called ``despair'' since the level 1 main effect reflects feelings of guilt and delusions, and the level 2 main effect reflects suicidal feelings and loss of interest in activities.

We note that the fitted model features seven conjunctive items where all attributes at particular levels are required in order to increase the response probability: hmnin, hpanx, happt, hsanx, hhypc, hintr, and hagit. Other items have more complex structure including main effects for some attributes, but only interactions for other pairs of attributes.

Table \ref{tab:appliclambda} shows the model's estimate of \(\lambda\), the matrix of coefficients relating the covariates age and sex to the latent state, as well as the estimates for \(\gamma_{k2}\) for all \(k \in [K]\), the thresholds for the augmented variable of each dimension of the multivariate probit specification. We see that for the set of respondents in this dataset, both being female and being older are positively associated with higher levels of attribute 1, ``anxiety.'' Both being female and being older are negatively associated with higher levels of attribute 2, ``weight loss.'' Being female is negatively associated with higher levels of ``despair'' (the coefficient for age for attribute 3 falls into the 95\% equal-tailed credible interval containing zero and thus is treated as not relevant).

The model's estimate of \(R\), the polychoric correlation matrix for the latent state, is displayed in Table \ref{tab:applicRmat}. We observe that there is a moderate amount of correlation between attributes 1 and 3.

Regarding the classification results, we observe that six latent state values comprise about 50\% of the dataset (see Table \ref{tab:appliclassific}). We write Low for an attribute level of 0, Medium for an attribute level of 1, and High for attribute value of 2. Four of these latent states have the highest anxiety level and two have the middle level of anxiety; all of these latent states have either the lowest or middle level of despair. For context, we note that the average age of all respondents was 40.972 years and that 62.677\% of the respondents were female. We also note that one can use the coefficients and threshold shown in Table \ref{tab:appliclambda} as well as the three fixed threshold values to see how different values of covariates relate to the latent state through the multivariate probit specification.

\begin{table}
\renewcommand\thetable{11}
\small
\begin{center}
\caption{\label{tab:appliclassific} Most common classifications}
\vspace{0.5\baselineskip}
\begin{tabular}{ccc|ccc}
\toprule
Anxiety & Weight loss & Despair & Proportion & Avg. age (years) & \% female \\
\midrule
High & Low & Medium & 0.111 & 47.582 & 0.642 \\
High & Medium & Low & 0.093 & 48.243 & 0.664 \\
Medium & Medium & Medium & 0.081 & 37.962 & 0.632 \\
High & Low & Low & 0.077 & 51.864 & 0.647 \\
Medium & High & Low & 0.067 & 38.663 & 0.543 \\
High & Medium & Medium & 0.062 & 43.917 & 0.675 \\
\bottomrule
\end{tabular}
\end{center}
\end{table}

\section{Discussion}

The model described in this paper represents an extension of the RLCM framework to include respondent-specific covariates in a cross-sectional model, while allowing for correlation between the attributes of the multi-leveled latent state of respondents using a multivariate probit specification. Here our model is illustrated for use with the Hamilton Rating Scale for Depression as an alternative to latent trait methods such as item response theory or factor analysis. Factor analysis and item response theory tend to utilize a linear response function with continuous latent traits, and are helpful when scores on each dimension are desired. The RLCM can more easily utilize interactions and study more complex response structures, and differs by resulting in a classification rather than a score. This can be helpful when there is a need to bin respondents according to some diagnosis. This exploratory model has a similar limitation to exploratory factor analysis, namely that care must be taken when making labels for interpretation of attributes. This can be even more challenging in this model that includes different levels of interactions rather than merely a factor loading matrix of main effects. However, all diagnostic models for polytomous data require careful consideration when labelling attributes. These models provide greater insight about the process whereby the attributes relate to observed responses. With that additional specificity comes the potential challenge for intereptation, and that motivates those who utilize these models to design instruments with clearly defined attributes.

Furthermore, when performing model selection, practitioners may want to allow their choice of model to depend not only on the procedure we outline above (running models of increasing complexity with posterior predictive checks), but also on previous results from applying similar models as well as past experience regarding the specific area of application. We note that in real world data, it is impossible to know for a particular number of latent attributes and levels whether or not there are latent classes which contain no respondents. However, reasonable indices for model selection would not select a model that has empty classes, since that would add to model complexity but not improve model fit.

Because many medical applications seek to classify subjects according to groups for which different interventions may be appropriate, RLCMs are a natural choice for achieving this aim. Here we illustrate the approach in a cross-sectional study, though in practice, subjects can be expected to move between latent classes which suggests that longitudinal extensions of this model could be valuable. In applications, practitioners should consider whether their interest is in classifying or scoring when choosing a modelling approach. If classification is the objective, RLCMs and the software to fit them are becoming more accessible and should be considered as alternatives to latent trait models.

\section*{Appendix}

\subsection*{Proof regarding monotonicity condition}\label{sec:proofmonoc}

In this subsection, we write \(d_u\) to mean the design vector for latent state \(u\), \(d_{hu}\) to be the \(h\)th component of \(d_u\), and \(d_{(a, b, c)u}\) to mean \(d_u\) less the components of \(d_u\) whose indices are \(a\), \(b\) and \(c\).

Regarding the region \(\mathcal{R}_j\), consider the monotonicity condition \eqref{monotoncond}. Whenever \(u \geq v\), there are three possibilities for a given position \(h\) in the vectors \(d_u\) and \(d_v\): (i) \(d_{hu} = 0 \,\land\, d_{hv} = 0\), (ii) \(d_{hu} = 1 \,\land\, d_{hv} = 0\), (iii) \(d_{hu} = 1 \,\land\, d_{hv} = 1\).

Note however that for \(h = 1\), for all \(\alpha \in A_L\), \(d_{h\alpha} = 1\). Therefore the condition is logically equivalent to \begin{equation}
\forall\, u, v \in A_L \quad u \geq v \implies d_{(1)u} \beta_{(1)j} \geq d_{(1)v} \beta_{(1)j}\end{equation}

In other words, the intercept element of \(\beta_j\) has no effect on the monotonicity condition. In what follows we use this form of the monotonicity condition.

Now consider the monotonicity condition in the following cases, where \(h > 1\):

Case 1: When \(\beta_{hj} = 0\), \begin{align}
& \forall\, u, v \in A_L \quad u \geq v \implies d_{(1)u} \beta_{(1)j} \geq d_{(1)v} \beta_{(1)j} \notag\\
\enspace\equiv\quad & \forall\, u, v \in A_L \quad u \geq v \implies d_{(1,h)u} \beta_{(1,h)j} \geq d_{(1,h)v} \beta_{(1,h)j} \notag\\
\enspace\equiv\quad & \forall\, u, v \in A_L \quad u \geq v \implies d_{(1,h)u} \beta_{(1,h)j} - d_{(1,h)v} \beta_{(1,h)j}\geq 0 \notag\\
\equiv\quad & \min_{u, v:\, u, v \in A_L \,\land\, u \geq v} \left(d_{(1,h)u} - d_{(1,h)v}\right) \beta_{(1,h)j} \geq 0 \notag\\
\equiv\quad & \max_{u, v:\, u, v \in A_L \,\land\, u \geq v} -\left(d_{(1,h)u} - d_{(1,h)v}\right) \beta_{(1,h)j} \leq 0 \label{monotoncondcase1}
\end{align}

We denote \eqref{monotoncondcase1} as \(\text{cond1}(\beta_{(1,h)j})\).

Case 2: \(\beta_{jh} \neq 0\): Consider the three possible situations for \(d_{hu}\) and \(d_{hv}\) mentioned above. In cases (i) and (iii), \(\beta_{hj}\) cancels from both sides and does not affect the condition. In case (ii), \(\beta_{hj}\) is on the left-hand side but not the right. For all \(h > 1\), all three cases occur.

Note now that no matter what value \(\beta_{hj}\) takes, cases (i) and (iii) do not affect the monotonicity condition. We consider case (ii). Observe that \begin{align}
& \forall\, u, v \in A_L \quad u \geq v \implies d_{(1)u} \beta_{(1)j} \geq d_{(1)v} \beta_{(1)j} \notag\\
\equiv\quad & \forall\, u, v \in A_L \quad u \geq v \implies \beta_{hj} \geq -\left(d_{(1,h)u} - d_{(1,h)v}\right) \beta_{(1,h)j} \notag\\
\equiv\quad & \beta_{hj} \geq \max_{u, v: u, v \in A_L \,\land\, u \geq v} -\left(d_{(1,h)u} - d_{(1,h)v}\right) \beta_{(1,h)j} =: L_{hj} \label{monotoncondcase2}
\end{align}

We denote \eqref{monotoncondcase2} as \(\text{cond2}(\beta_{(1)j})\).

We observe that under our model: \[\begin{array}{ll}
\text{if } \delta_{hj} = 0, \text{then } & \beta_j \in \mathcal{R}_j \iff \beta_{hj} = 0 \,\land\, \text{cond1}(\beta_{(1,h)j}) \\
\text{if } \delta_{hj} = 1, \text{then } & \beta_j \in \mathcal{R}_j \iff \text{cond2}(\beta_{(1)j})
\end{array}\]

\noindent (For the second of these, note that under the slab, the probability that \(\beta_{hj} = 0\) is zero, so Case 1 and thus \(\text{cond1}(\beta_{(1,h)j})\) is negligible.)

\section*{}

\textbf{Data availability} \quad Data and/or research tools used in the preparation of this manuscript were obtained from the National Institute of Mental Health (NIMH) Data Archive (NDA). NDA is a collaborative informatics system created by the National Institutes of Health to provide a national resource to support and accelerate research in mental health. Dataset identifier(s): \url{https://dx.doi.org/10.15154/as7z-bc83}. This manuscript reflects the views of the authors and may not reflect the opinions or views of the NIH or of the Submitters submitting original data to NDA.

\vspace{\baselineskip}

\noindent \textbf{Code availability} \quad A Python package, \texttt{probitlcm}, containing an implementation of the procedures for simulation, data analysis, and model selection described in this manuscript has been released under a FLO (free/libre/open) license and is available at \url{https://github.com/ericwayman01/probitlcm}. The parameter values used in the simulations are available at \url{https://github.com/ericwayman01/probitlcm_params}.

\vspace{\baselineskip}

\noindent \textbf{Funding} \quad This work was partially supported by the U.S. National Science Foundation, under award numbers SES 2150628 and SES 1951057.

\section*{Ethics declarations}

\textbf{Conflicts of interest} \quad On behalf of all authors, the corresponding author states that there is no conflict of interest.

\bibliographystyle{plainnat}
\bibliography{refs}

\newpage

\noindent {\textbf{\LARGE Supplementary Materials}}

\section*{Supplementary Material A}

Supplementary Material A describes the transformation and the transformed model.

\subsection*{Defining the transformations}

We write \(z = (z_1, \ldots, z_8)\) where \begin{equation}
  \begin{array}{lll}
    z_1 & = & Y \\
    z_2 & = & \alpha \\
    z_3 & = & b_1 = (Y^\ast, \beta, \delta, \omega) \\
    z_4 & = & \alpha^\ast \\
    z_5 & = & \gamma \\
    z_6 & = & \lambda \\
    z_7 & = & (r_{12}, r_{13}, \ldots, r_{1K}, r_{23}, \ldots, r_{2K}, \ldots, r_{K-1,K}) \\
    z_8 & = & (v_1, \ldots, v_K). \\
  \end{array}
\end{equation}

By the theorem regarding change of variables \citep{casella2002statistical,marden2017mathematical}, we have \(p_{\widetilde{Z}}(\widetilde{z}) = p_{Z}(g^{-1}(\widetilde{z})) \cdot \left|\det{J_{g^{-1}}(\widetilde{z})}\right|\) where \(J_{g^{-1}}(\widetilde{z})\) denotes the Jacobian matrix of \(g^{-1}\) evaluated at \(\widetilde{z}\).

We have \(V = \text{diag}(z_8)\). We define \(g(z) = (g_1(z), \ldots, g_8(z))\), where \begin{equation}
  \begin{array}{lll}
    g_1(z) & = & Y \\
    g_2(z) & = & \alpha \\
    g_3(z) & = & b_1 \\
    g_4(z) & = & \alpha^{\ast} V^{1/2} \\
    g_5(z) & = & \gamma V^{1/2} \\
    g_6(z) & = & \lambda V^{1/2} \\
    g_7(z) & = & (v_1^{1/2} r_{12} v_2^{1/2}, v_1^{1/2} r_{13} v_3^{1/2}, \ldots, v_1^{1/2} r_{1K} v_K^{1/2}, v_2^{1/2} r_{23} v_3^{1/2}, \ldots, v_2^{1/2} r_{2K} v_K^{1/2}, \\
    & & \quad\quad \ldots, v_{K-1}^{1/2} r_{K-1,K} v_K^{1/2}) \\
    g_8(z) & = & (v_1, \ldots, v_K). \\
  \end{array}
 \end{equation}

\noindent The value taken by \(g\) at \(z\) is \(\widetilde{z}\). We write \(\widetilde{z} = (\widetilde{z}_1, \ldots, \widetilde{z}_8)\) where \begin{equation}
  \begin{array}{lll}
    \widetilde{z}_1 & = & Y \\
    \widetilde{z}_2 & = & \alpha \\
    \widetilde{z}_3 & = & b_1 \\
    \widetilde{z}_4 & = & \widetilde{\alpha^\ast} \\
    \widetilde{z}_5 & = & \widetilde{\gamma} \\
    \widetilde{z}_6 & = & \widetilde{\lambda} \\
    \widetilde{z}_7 & = & (\sigma_{12}, \ldots, \sigma_{1K}, \sigma_{23}, \ldots, \sigma_{2K}, \ldots, \sigma_{K-1,K}) \\
    \widetilde{z}_8 & = & (\sigma_{11}, \ldots, \sigma_{KK}). \\  
  \end{array}
\end{equation}

We now consider \(g^{-1}\). The value taken by \(g^{-1}\) at \(\widetilde{z}\) is \(z\). We write \(z = g^{-1}(\widetilde{z}) = (h_1(\widetilde{z}), \ldots, h_8(\widetilde{z}))\), where, writing \(\widetilde{V} = \text{diag}(\sigma_{11}, \ldots, \sigma_{KK})\),

\begin{equation}
  \begin{array}{lll}
    h_1(\widetilde{z}) & = & Y \\
    h_2(\widetilde{z}) & = & \alpha \\
    h_3(\widetilde{z}) & = & b_1 \\
    h_4(\widetilde{z}) & = & \widetilde{\alpha^\ast} \widetilde{V}^{-1/2} \\
    h_5(\widetilde{z}) & = & \widetilde{\gamma} \widetilde{V}^{-1/2} \\
    h_6(\widetilde{z}) & = & \widetilde{\lambda} \widetilde{V}^{-1/2} \\
    h_7(\widetilde{z}) & = & \left(\frac{\sigma_{12}}{\sigma_{11}^{1/2} \sigma_{22}^{1/2}}, \ldots, \frac{\sigma_{1K}}{\sigma_{11}^{1/2} \sigma_{KK}^{1/2}}, \frac{\sigma_{23}}{\sigma_{22}^{1/2} \sigma_{33}^{1/2}}, \ldots \frac{\sigma_{2K}}{\sigma_{22}^{1/2} \sigma_{KK}^{1/2}}, \cdots, \frac{\sigma_{K-1,K}}{\sigma_{K-1,K-1}^{1/2} \sigma_{KK}^{1/2}}\right) \\
    h_8(\widetilde{z}) & = & (\sigma_{11}, \ldots, \sigma_{KK}).
  \end{array}
\end{equation}

The Jacobian matrix of \(g^{-1}\) evaluated at \(\widetilde{z}\) \cite[page 66]{marden2017mathematical} is \begin{equation*}
  (J_{g^{-1}}(\widetilde{z}))_{ij} := \frac{\partial}{\partial \widetilde{z}_j} h_i(\widetilde{z}_j)
\end{equation*}

\noindent which can be written as a \(8 \times 8\) block matrix, itself consisting of four superblocks: \begin{equation*}
  \begin{pmatrix}
    B_{11} & B_{12} \\
    B_{21} & B_{22}
  \end{pmatrix}
\end{equation*}

\noindent where \begin{equation*}
  B_{11} = \begin{pmatrix}
    I & O & O \\
    O & I & O \\
    O & O & I \\
    \end{pmatrix},
\end{equation*}

\noindent \(B_{12} = O\), and \(B_{21} = O\). We claim that \(J_{g^{-1}}(\widetilde{z})\) is an upper-triangular matrix. For this to be true, it remains to be shown that \(B_{22}\) is upper-triangular. \(B_{22}\) itself is a \(5 \times 5\) block matrix. Let \((B_{22})_{ij}\) denote the block in position \(ij\) of \(B_{22}\).

Observe that \((B_{22})_{11} = \partial h_4(\widetilde{z}) / \partial \widetilde{z}_4 = \widetilde{\alpha^\ast} \widetilde{V}^{-1/2} / \partial \widetilde{\alpha^\ast}\), and that all the blocks below \((B_{22})_{11}\) evaluate to \(O\).

Observe that \((B_{22})_{22} = \partial h_5(\widetilde{z}) / \partial \widetilde{z}_5 = \partial \widetilde{\gamma} \widetilde{V}^{-1/2} / \partial \widetilde{\gamma}\), and that all the blocks below \((B_{22})_{22}\) evaluate to \(O\).

Observe that \((B_{22})_{33} = \partial h_6(\widetilde{z}) / \partial \widetilde{z}_6 = \partial \widetilde{\lambda} \widetilde{V}^{-1/2} / \partial \widetilde{\lambda}\), and that all the blocks below \((B_{22})_{33}\) evaluate to \(O\).

Observe that \begin{align*}
  (B_{22})_{44} & = \partial h_7(\widetilde{z}) / \partial \widetilde{z}_7 \\
  & = \text{diag}\left(\frac{1}{\sigma_{11}^{1/2} \sigma_{22}^{1/2}}, \ldots, \frac{1}{\sigma_{11}^{1/2} \sigma_{KK}^{1/2}}, \frac{1}{\sigma_{22}^{1/2} \sigma_{33}^{1/2}}, \ldots \frac{1}{\sigma_{22}^{1/2} \sigma_{KK}^{1/2}}, \cdots, \frac{1}{\sigma_{K-1,K-1}^{1/2} \sigma_{KK}^{1/2}}\right)
\end{align*}

\noindent and, evaluating \((B_{22})_{54}\) (the block below it) gives \begin{equation*}
  (B_{22})_{54} = \partial h_8(\widetilde{z}) / \partial \widetilde{z}_7 = O.
\end{equation*}

Finally, observe that \((B_{22})_{55} = \partial h_8(\widetilde{z}) / \partial \widetilde{z}_8 = I\).

We conclude that \(J_{g^{-1}}(\widetilde{z})\) is an upper-triangular block matrix. Therefore the determinant of \(J_{g^{-1}}(\widetilde{z})\) is the product of the determinants of the diagonal blocks \citep[page 25]{horn1990matrix}. Specifically, \begin{align}
  & \det{J_{g^{-1}}(\widetilde{z})} \notag\\
  & = \det\left(\frac{\partial}{\partial \widetilde{\alpha^\ast}} \widetilde{\alpha^\ast} \widetilde{V}^{-1/2}\right) \cdot \det\left(\frac{\partial}{\partial \widetilde{\gamma}} \widetilde{\gamma} \widetilde{V}^{-1/2}\right) \cdot \det\left(\frac{\partial}{\partial \widetilde{\lambda}} \widetilde{\lambda} \widetilde{V}^{-1/2}\right) \cdot \det\left(\frac{\partial}{\partial \widetilde{z}_7} \partial h_7(\widetilde{z})\right) \notag\\
  & = \underbrace{\left(\prod_{k \in [K]} \sigma_{kk}^{-1/2}\right)^N}_{(\text{D1})} \cdot \underbrace{\left(\prod_{k \in [K]} \sigma_{kk}^{-1/2}\right)^{L - 2}}_{(\text{D2})} \cdot \underbrace{\left(\prod_{k \in [K]} \sigma_{kk}^{-1/2}\right)^{D}}_{(\text{D3})} \cdot \underbrace{\left(\prod_{k \in [K]} \sigma_{kk}^{-1/2}\right)^{K - 1}}_{(\text{D4})}.
\end{align}

Observe that since all the terms are positive, we have \(|\det{J_{g^{-1}}(\widetilde{z})}| = \det{J_{g^{-1}}(\widetilde{z})}\).

\subsection*{Some observations}

In this section, we are referring to the realized values of random variables.

Note that \(R\) is the matrix formulation of \(z_7\) with many duplicate elements. When we write phrases like ``the density of \(R\)'' we ignore these duplicate elements. Similarly, \(\Sigma\) is the matrix formulation of \(\widetilde{z}_7\).

\(h_7(\widetilde{z})\) is a vector containing the unique elements of \(R\) written in terms of elements of \(\widetilde{z}\). \(g_7(z)\) is a vector containing the unique elements of \(\Sigma\) written in terms of elements of \(z\).

We note that component by component, each element of the vector \(g_8(z)\) equals the corresponding element of the vector \(\widetilde{z}_8\), so this component of the transformation is just a relabeling.

\subsection*{Finding the density for the transformed model}

From the text of the manuscript, we have that the density function for \(Z\), namely \(p_Z(z)\), using the shorthand notation of the main manuscript, is \begin{align}
    p(Z) & = \underbrace{p(Y \mid Y^\ast, \kappa) \cdot p(Y^\ast \mid \beta, \alpha) \cdot p(\kappa) \cdot p(\beta \mid \delta) \cdot p(\delta \mid \omega) \cdot p(\omega)}_{(\text{part1})} \notag\\
& \quad\quad \cdot \underbrace{p(\alpha \mid \alpha^\ast, \gamma)}_{(\text{part2})} \cdot \underbrace{p(\gamma \mid V)}_{(\text{part3})} \cdot \underbrace{p(R, V)}_{(\text{part4})} \cdot \underbrace{p(\alpha^\ast \mid R, \lambda)}_{(\text{part5})} \cdot \underbrace{p(\lambda \mid R)}_{(\text{part6})}
\end{align}

\noindent a product of density functions, where we have assigned a label to each part. Utilizing \(p_{\widetilde{Z}}(\widetilde{z}) = p_{Z}(g^{-1}(\widetilde{z})) \cdot \left|\det{J_{g^{-1}}(\widetilde{z})}\right|\), we have that \begin{align}
  & p_{\widetilde{Z}}(\widetilde{z}) = (\widetilde{\text{part1}}) \cdot (\widetilde{\text{part2}}) \cdot (\widetilde{\text{part3}}) \cdot (\widetilde{\text{part4}}) \cdot (\widetilde{\text{part5}}) \cdot (\widetilde{\text{part6}})
\end{align}

\noindent where each of \((\widetilde{\text{part1}}), \ldots, (\widetilde{\text{part6}})\) is written out in full in the following.

\subsubsection*{Transformation of part1}

\begin{align}
  (\widetilde{\text{part1}}) & = (\text{part1}) \notag\\
  & = \prod_{n=1}^N \left[\prod_{j=1}^J I(Y_{nj}^\ast \in (\kappa_{j, Y_{nj} - 1}, \kappa_{jY_{nj}}]) \cdot \phi(Y_{nj}^\ast; d_n \beta_j, 1)\right] \notag\\
    & \quad\quad \cdot I(-\infty = \kappa_{j0} < 0 = \kappa_{j1} < \cdots < \kappa_{j Mj} = \infty) \notag\\
    & \quad\quad \cdot \prod_{j=1}^J \Bigg[c_j(\delta_j) \cdot I(\beta_j \in \mathcal{R}_j) \notag\\
      &\quad\quad \cdot \left(\prod_{h=1}^H \left[I(\delta_{hj} = 0) \cdot \Delta(\beta_{hj}) + I(\delta_{hj} = 1) \cdot \phi(\beta_{hj}; 0, \sigma_{\beta}^2)\right]\right) \notag\\
      & \quad\quad \cdot \left(\prod_{h=1}^H \omega^{\delta_{hj}} (1 - \omega)^{1 - \delta_{hj}}\right)\Bigg] \cdot \frac{1}{B(\omega_0, \omega_1)} \omega^{\omega_0 - 1} (1 - \omega)^{\omega_1 - 1}.
\end{align}

\subsubsection*{Transformation of part2}

\begin{align}
  (\widetilde{\text{part2}}) & = \prod_{n=1}^N \prod_{k=1}^K I\left(\widetilde{\alpha^\ast}_{nk} \sigma_{kk}^{-1/2} \in \Big(\widetilde{\gamma}_{k\alpha_{nk}} \sigma_{kk}^{-1/2}, \widetilde{\gamma}_{k,\alpha_{nk}+1} \sigma_{kk}^{-1/2}\Big]\right) \notag\\
    & = \prod_{n=1}^N \prod_{k=1}^K I\left(\widetilde{\alpha^\ast}_{nk} \in \Big(\widetilde{\gamma}_{k \alpha_{nk}}, \widetilde{\gamma}_{k,\alpha_{nk}+1}\Big]\right).
\end{align}

\subsubsection*{Transformation of part3}

\begin{align}
  & (\widetilde{\text{part3}}) \notag\\
  & = \prod_{l=2}^{L - 1} \Bigg[a \sigma_{kk}^{1/2} \exp\left[-a \sigma_{kk}^{1/2} \cdot (\widetilde{\gamma}_{kl} \sigma_{kk}^{-1/2} - \widetilde{\gamma}_{k,l-1} \sigma_{kk}^{-1/2})\right] \notag\\
& \quad\quad \cdot I\left(\widetilde{\gamma}_{kl} \sigma_{kk}^{-1/2} \in (\widetilde{\gamma}_{k, l-1} \sigma_{kk}^{-1/2}, \infty)\right)\Bigg] \cdot (\text{D2}) \notag\\
  & = \prod_{l=2}^{L - 1} \Bigg[a \sigma_{kk}^{1/2} \exp\left[-a (\widetilde{\gamma}_{kl} - \widetilde{\gamma}_{k,l-1})\right] \notag\\
  & \quad\quad \cdot I\left(\widetilde{\gamma}_{kl} \sigma_{kk}^{-1/2} \in (\widetilde{\gamma}_{k, l-1} \sigma_{kk}^{-1/2}, \infty)\right)\Bigg] \cdot (\text{D2}) \notag\\
  & = \prod_{l=2}^{L - 1} \Bigg[a \sigma_{kk}^{1/2} \exp\left[-a (\widetilde{\gamma}_{kl} - \widetilde{\gamma}_{k,l-1})\right] \cdot I\left(\widetilde{\gamma}_{kl} \in (\widetilde{\gamma}_{k, l-1}, \infty)\right)\Bigg] \cdot (\text{D2}) \notag\\
& = \prod_{l=2}^{L - 1} \Bigg[a \exp\left[-a (\widetilde{\gamma}_{kl} - \widetilde{\gamma}_{k,l-1})\right] \cdot I\left(\widetilde{\gamma}_{kl} \in (\widetilde{\gamma}_{k, l-1}, \infty)\right)\Bigg].
\end{align}

\noindent which is the product of densities of left-truncated exponentials.

\subsubsection*{Transformation of part4}

Observing the following algebraic fact: \begin{equation}
  \sum_{k=1}^K v_k^{-1} A_{kk} = \text{tr}(V^{-1} R^{-1}) = \text{tr}(V^{-1/2} R^{-1} V^{-1/2}) = \text{tr}[(V^{1/2} R V^{1/2})^{-1}].\label{part4pt1}
\end{equation}

\noindent Define the matrix \(S\) by \begin{equation*}
  (S)_{ij} = \begin{pmatrix}\frac{\sigma_{ij}}{\sigma_{ii}^{1/2} \sigma_{jj}^{1/2}}\end{pmatrix}.
\end{equation*}

\noindent Observe that \(\widetilde{V}^{1/2} S \widetilde{V}^{1/2} = \Sigma\). Thus, under the transformation, replacing the appropriate quantities, \begin{align*}
  \eqref{part4pt1} = \text{tr}[(\widetilde{V}^{1/2} S \widetilde{V}^{1/2})^{-1}] = \text{tr}[\Sigma^{-1}].
\end{align*}

\noindent In addition, observe that \begin{align}
  \prod_{k \in [K]} \sigma_{kk}^{-\frac{1}{2}(v_0 + 2)} \cdot (\text{D4}) & = \prod_{k \in [K]} \sigma_{kk}^{-\frac{1}{2}(v_0 + 2)} \cdot \prod_{k \in [K]} \sigma_{kk}^{-\frac{1}{2}(K - 1)} \notag\\
  & = \prod_{k \in [K]} \sigma_{kk}^{-\frac{1}{2}(v_0 + K + 1)} \notag\\
  & = (\det{\widetilde{V}})^{-\frac{1}{2}(v_0 + K + 1)}.
\end{align}

\noindent Note that \begin{align}
  & (\det{S})^{-\frac{1}{2}(v_0 + K + 1)} \cdot (\det{\widetilde{V}})^{-\frac{1}{2}(v_0 + K + 1)} \notag\\
  & = [(\det{\widetilde{V}})^{1/2} (\det{S}) (\det{\widetilde{V}})^{1/2}]^{-\frac{1}{2}(v_0 + K + 1)} \notag\\
  & = [\det(\Sigma)]^{-\frac{1}{2}(v_0 + K + 1)}
\end{align}

\noindent and thus by the change of variables, we have \begin{align}
  (\widetilde{\text{part4}}) & = c_1 \cdot [(\det{\widetilde{V}})^{1/2} (\det{S}) (\det{\widetilde{V}})^{1/2}]^{-\frac{1}{2}(v_0 + K + 1)} \notag\\
  & \quad\quad \cdot \prod_{k \in [K]} \left[\exp\left(-\frac{1}{2} \text{tr}[(\widetilde{V}^{1/2} S \widetilde{V}^{1/2})^{-1}]\right)\right] \notag\\
  & = (\det{\Sigma})^{-\frac{1}{2}(v_0 + K + 1)} \exp\left(-\frac{1}{2}\tr(\Sigma^{-1})\right) \label{part4}
\end{align}

\noindent which is proportional to the density of an Inverse Wishart.

\subsubsection*{Transformation of part5}

Writing \(\text{etr}(\cdot)\) to mean \(\exp(\text{tr}(\cdot))\), \begin{align}
  & (\widetilde{\text{part5}}) \notag\\
  & = (2\pi)^{-\frac{1}{2}KN} (\det{S})^{-\frac{1}{2}N} \notag\\
  & \quad\quad \cdot \text{etr}\left\{-\frac{1}{2}\left(\widetilde{\alpha^\ast} \widetilde{V}^{-1/2} - X \widetilde{\lambda} \widetilde{V}^{-1/2}\right) (S)^{-1} \left(\widetilde{\alpha^\ast} \widetilde{V}^{-1/2} - X \widetilde{\lambda} \widetilde{V}^{-1/2}\right)^\prime\right\} \cdot (\text{D1}) \notag\\
  & = (2\pi)^{-\frac{1}{2}KN} (\det{S})^{-\frac{1}{2}N} \text{etr}\left\{-\frac{1}{2}\left(\widetilde{\alpha^\ast} - X \widetilde{\lambda}\right) \Sigma^{-1} \left(\widetilde{\alpha^\ast} - X \widetilde{\lambda}\right)^\prime\right\} \cdot (\det{\widetilde{V}})^{-\frac{1}{2}N} \notag\\
  & = (2\pi)^{-\frac{1}{2}KN} (\det{\Sigma})^{-\frac{1}{2}N} \text{etr}\left\{-\frac{1}{2}\left(\widetilde{\alpha^\ast} - X \widetilde{\lambda}\right) \Sigma^{-1} \left(\widetilde{\alpha^\ast} - X \widetilde{\lambda}\right)^\prime\right\} \label{part5}
\end{align}

\noindent which is the density of a matrix variate normal (see Supplementary Material C).

\subsubsection*{Transformation of part6}

\begin{align}
  & (\widetilde{\text{part6}}) \notag\\
  & = (2\pi)^{-\frac{1}{2}DK} (\det{S})^{-\frac{1}{2}D} (\det{I_D})^{-\frac{1}{2}K} \cdot \text{etr}\left\{-\frac{1}{2} \left(\widetilde{\lambda} \widetilde{V}^{-1/2}\right) S^{-1} \left(\widetilde{\lambda} \widetilde{V}^{-1/2}\right)^\prime\right\} \cdot (\text{D3}) \notag\\
  & = (2\pi)^{-\frac{1}{2}DK} (\det{S})^{-\frac{1}{2}D} (\det{I_D})^{-\frac{1}{2}K} \cdot \text{etr}\left\{-\frac{1}{2} \widetilde{\lambda} \Sigma^{-1} \widetilde{\lambda}^\prime\right\} \cdot (\det{\widetilde{V}})^{-\frac{1}{2}D} \notag\\
  & = (2\pi)^{-\frac{1}{2}DK} (\det{\Sigma})^{-\frac{1}{2}D} (\det{I_D})^{-\frac{1}{2}K} \cdot \text{etr}\left\{-\frac{1}{2} \widetilde{\lambda} \Sigma^{-1} \widetilde{\lambda}^\prime\right\}
\end{align}

\noindent which is proportional to the density of a matrix variate normal.

\newpage

\section*{Supplementary Material B}

Supplementary Material B contains derivations of conditional distributions used for sampling various parameters.

\subsection*{Latent states and related auxiliary variables}

For each \(k \in [K]\) and \(n \in [N]\), we sample \((\widetilde{\alpha^{\ast}}_{nk}, \alpha_{nk})\) by first sampling \(\alpha_{nk}\) with a Gibbs step collapsing on \(\widetilde{\alpha^{\ast}}_{nk}\), and then using that value to sample \(\widetilde{\alpha^{\ast}}_{nk}\) from its full conditional. To find the density for that first step, we integrate the full conditional of \((\widetilde{\alpha^{\ast}}_{nk}, \alpha_{nk})\) with respect to \(\widetilde{\alpha^{\ast}}_{nk}\); that full conditional is proportional to \begin{equation}\label{bg3part1}
  p(\widetilde{\alpha^{\ast}}_{nk}, \alpha_{nk} \mid \alpha_{n(k)}, \widetilde{\gamma}_k, Y_{nj}^\ast, \beta_j) = c_1 \cdot (\widetilde{\text{part1}}) \cdot (\widetilde{\text{part2}}) \cdot (\widetilde{\text{part5}}).
\end{equation}

We observe that \((\widetilde{\text{part5}})\) is the density of a matrix variate normal with variable \(\widetilde{\alpha^\ast}\) with parameters \(X\widetilde{\lambda}\) and \(I_N \otimes \Sigma\). Using some algebra we have that \begin{equation}
  (\widetilde{\text{part5}}) = \prod_{i \in [N]} (2\pi)^{-\frac{1}{2}K} (\det{\Sigma})^{-\frac{1}{2}} \exp\left(-\frac{1}{2} (\widetilde{\alpha^{\ast}}_i - X_i \widetilde{\lambda}) \Sigma^{-1} (\widetilde{\alpha^{\ast}}_i - X_i \widetilde{\lambda})^\prime\right)
\end{equation}

\noindent the product of \(N\) densities of multivariate normals, each multivariate normal density having variable \(\widetilde{\alpha^\ast}_i\), mean \(X_i \widetilde{\lambda}\) and variance \(\Sigma\). Using \citet{marden2015multivariate} Proposition 3.3 we have that certainly for \(n\) that the density of \(\widetilde{\alpha^\ast}_n\) can be written as the product of two densities, one being a function only of \(\widetilde{\alpha^\ast}_{n(k)}\), and the other being a normal density with variable \(\widetilde{\alpha^\ast}_{nk}\), mean \(\mu_{nk} = x_n \lambda_k + (\widetilde{\alpha^{\ast}}_{n(k)} - x_n \lambda_{(k)})\Sigma_{(k)(k)}^{-1} \Sigma_{(k)k}\), and variance \(\sigma_k^2 = \Sigma_{kk} - \Sigma_{k(k)} \Sigma_{(k)(k)}^{-1} \Sigma_{(k)k}\).

Using the above, we have that \begin{align}
  \eqref{bg3part1} & = c_2 \cdot \left[\prod_{j=1}^J \phi(Y_{nj}^\ast; d_n \beta_j, 1)\right] \cdot I\left(\widetilde{\alpha^\ast}_{nk} \in \Big(\widetilde{\gamma}_{k\alpha_{nk}}, \widetilde{\gamma}_{k,\alpha_{nk}+1}\Big]\right) \cdot \phi(\widetilde{\alpha^\ast}_{nk}; \mu_{nk}, \sigma_k^2).
\end{align}

\noindent Therefore \begin{align}
& p(\alpha_{nk} \mid \alpha_{n(k)}, \widetilde{\gamma}_k, Y_{nj}^\ast, \beta_j) \notag\\
  & = \int p(\widetilde{\alpha^{\ast}}_{nk}, \alpha_{nk} \mid \alpha_{n(k)}, \widetilde{\gamma}_k, Y_{nj}^\ast, \beta_j) d\widetilde{\alpha^{\ast}}_{nk} \notag\\
  & = c_2 \cdot \left[\prod_{j=1}^J \phi(Y_{nj}^\ast; d_n \beta_j, 1)\right] \cdot \int I\left(\widetilde{\alpha_{nk}^\ast} \in \Big(\widetilde{\gamma}_{k\alpha_{nk}}, \widetilde{\gamma}_{k,\alpha_{nk}+1}\Big]\right) \cdot \phi(\widetilde{\alpha_{nk}^\ast}; \mu_{nk}, \sigma_k^2) d\widetilde{\alpha^{\ast}}_{nk} \notag\\
  & = c_2 \cdot \left[\prod_{j=1}^J \phi(Y_{nj}^\ast; d_n \beta_j, 1)\right] \cdot \int_{\widetilde{\gamma}_{k \alpha_{nk}}}^{\widetilde{\gamma}_{k, \alpha_{nk} + 1}} \phi(\widetilde{\alpha^{\ast}}_{nk};\, \mu_{nk}, \sigma_k^2)\, d\widetilde{\alpha^{\ast}}_{nk} \notag\\
  & = c_2 \cdot \left[\prod_{j=1}^J \phi(Y_{nj}^\ast; d_n \beta_j, 1)\right] \cdot \left[\Phi\left(\frac{\widetilde{\gamma}_{k, \alpha_{nk} + 1} - \mu_{nk}}{\sigma_k}\right) - \Phi\left(\frac{\widetilde{\gamma}_{k \alpha_{nk}} - \mu_{nk}}{\sigma_k}\right)\right].
\end{align}

We calculate the probabilities for each value \(l \in \{0, \ldots, L - 1\}\) as follows. We plug the value \(l\) for \(\alpha_{nk}\) and all the other known values into \begin{equation}
  p_l := \left[\prod_{j=1}^J \phi(Y_{nj}^\ast; d_n \beta_j, 1)\right] \cdot \left[\Phi\left(\frac{\widetilde{\gamma}_{k, \alpha_{nk} + 1} - \mu_{nk}}{\sigma_k}\right) - \Phi\left(\frac{\widetilde{\gamma}_{k \alpha_{nk}} - \mu_{nk}}{\sigma_k}\right)\right].
\end{equation}

\noindent After doing that for each \(l\), we then find \(c := \left(\sum_{l=0}^{L_1 - 1} p_l\right)^{-1}\), and we have the result that for all \(\alpha_{nk} \in \{0, \ldots, L - 1\}\)\begin{equation}
p(\alpha_{nk} \mid \alpha_{n(k)}, \widetilde{\gamma}_k, Y_{nj}^\ast, \beta_j) = c \cdot p_{\alpha_{nk}}.
\end{equation}

Now, using proportionality and the observations from the process of finding (b), we have that the full conditional of \(\widetilde{\alpha^{\ast}}_{nk}\) is \begin{align}
  p(\widetilde{\alpha^{\ast}}_{nk} \mid \widetilde{\lambda}, \Sigma, \alpha_{nk}, \widetilde{\gamma}_k) & = c_3 \cdot (\widetilde{\text{part2}}) \cdot (\widetilde{\text{part5}}) \notag\\
  & = c_3 \cdot I\left(\widetilde{\alpha^{\ast}}_{nk} \in (\widetilde{\gamma}_{k\alpha_{nk}}, \widetilde{\gamma}_{k,{\alpha_{nk} + 1}})\right) \cdot \phi(\widetilde{\alpha^\ast}_{nk}; \mu_{nk}, \sigma_k^2)
\end{align}

\noindent so we conclude that \begin{align}
  & p(\widetilde{\alpha^{\ast}}_{nk} \mid \widetilde{\lambda}, \Sigma, \alpha_{nk}, \widetilde{\gamma}_k) \notag\\
  & = I\left(\widetilde{\alpha^{\ast}}_{nk} \in (\widetilde{\gamma}_{k\alpha_{nk}}, \widetilde{\gamma}_{k,{\alpha_{nk} + 1}})\right) \frac{\phi(\widetilde{\alpha^{\ast}}_{nk}; \mu_{nk}, \sigma_k^2)}{\Phi(\widetilde{\gamma}_{k,\alpha_{nk} + 1}; \mu_{nk}, \sigma_k^2) - \Phi(\widetilde{\gamma}_{k\alpha_{nk}}; \mu_{nk}, \sigma_k^2)}
\end{align}

\noindent a truncated normal density with left and right truncation points \(\widetilde{\gamma}_{k\alpha_{nk}}\) and \(\widetilde{\gamma}_{k,\alpha_{nk} + 1}\) respectively and whose underlying normal random variable has mean \(\mu_{nk}\) and variance \(\sigma_k^2\).

\subsection*{Thresholds for latent state levels}

The full conditional for \(\widetilde{\gamma}_{kl}\) for \(k \in [K]\) depends on the value for \(l\). For \(l \in \{2, 3, \ldots, L - 2\}\), the full conditional is \begin{align}\label{bg4part1}
  & p(\widetilde{\gamma}_{kl} \mid \widetilde{\gamma}_{k,l-1}, \widetilde{\gamma}_{k,l+1}, \widetilde{\alpha^\ast}) \notag\\
  & = c_1 \cdot (\widetilde{\text{part2}}) \cdot (\widetilde{\text{part3}}) \notag\\
  & = c_2 \cdot \left[\prod_{n \in [N]: \alpha_{nk} = l-1} I\left(\widetilde{\alpha^\ast}_{nk} \in \Big(\widetilde{\gamma}_{k, l-1}, \widetilde{\gamma}_{kl}\Big]\right)\right] \cdot \left[\prod_{n \in [N]: \alpha_{nk} = l} I\left(\widetilde{\alpha^\ast}_{nk} \in \Big(\widetilde{\gamma}_{kl}, \widetilde{\gamma}_{k,l+1}\Big]\right)\right] \notag\\
  & \quad\quad \cdot \exp\left[-a (\widetilde{\gamma}_{k, l+1} - \widetilde{\gamma}_{kl})\right] \cdot I\left(\widetilde{\gamma}_{k, l+1} \in (\widetilde{\gamma}_{k, l}, \infty)\right) \notag\\
  & \quad\quad \cdot \exp\left[-a (\widetilde{\gamma}_{kl} - \widetilde{\gamma}_{k, l-1})\right] \cdot I\left(\widetilde{\gamma}_{kl} \in (\widetilde{\gamma}_{k, l-1}, \infty)\right) \notag\\
  & = c_3 \cdot \left[\prod_{n \in [N]:\, \alpha_{nk} = l - 1} I(\widetilde{\gamma}_{kl} \geq \widetilde{\alpha^{\ast}}_{nk}) \cdot I(\widetilde{\gamma}_{k,l-1} < \widetilde{\alpha^{\ast}}_{nk})\right] \notag\\
  & \quad\quad \cdot \left[\prod_{n \in [N]:\, \alpha_{nk} = l} I(\widetilde{\gamma}_{k,l+1} \geq \widetilde{\alpha^{\ast}}_{nk}) \cdot I(\widetilde{\gamma}_{kl} < \widetilde{\alpha^{\ast}}_{nk})\right] \cdot I\left(\widetilde{\gamma}_{kl} \in (\widetilde{\gamma}_{k, l-1}, \widetilde{\gamma}_{k, l+1})\right) \notag\\
  & = c_4 \cdot \left[\prod_{n \in [N]:\, \alpha_{nk} = l - 1} I(\widetilde{\gamma}_{kl} \geq \widetilde{\alpha^{\ast}}_{nk})\right] \cdot \left[\prod_{n \in [N]:\, \alpha_{nk} = l} I(\widetilde{\gamma}_{kl} < \widetilde{\alpha^{\ast}}_{nk})\right] \notag\\
  & \quad\quad \cdot I\left(\widetilde{\gamma}_{kl} \in (\widetilde{\gamma}_{k, l-1}, \widetilde{\gamma}_{k, l+1})\right) \notag\\
  & = c_4 \cdot I\left(\widetilde{\gamma}_{kl} \geq \max_{n \in [N]:\, \alpha_{nk} = l - 1} \left(\widetilde{\alpha^{\ast}}_{nk}\right)\right) \cdot I\left(\widetilde{\gamma}_{kl} < \min_{n \in [N]:\, \alpha_{nk} = l} \left(\widetilde{\alpha^{\ast}}_{nk}\right)\right) \notag\\
  & \quad\quad \cdot I\left(\widetilde{\gamma}_{kl} \in (\widetilde{\gamma}_{k, l-1}, \widetilde{\gamma}_{k, l+1})\right) \notag\\
  & = c_4 \cdot I\left(\widetilde{\gamma}_{kl} \geq \max\left(\max_{n \in [N]:\, \alpha_{nk} = l - 1} \left(\widetilde{\alpha^{\ast}}_{nk}\right), \widetilde{\gamma}_{k,l-1}\right)\right) \notag\\
  & \quad\quad \cdot I\left(\widetilde{\gamma}_{kl} < \min\left(\min_{n \in [N]:\, \alpha_{nk} = l} \left(\widetilde{\alpha^{\ast}}_{nk}\right), \widetilde{\gamma}_{k,l+1}\right)\right)
\end{align}

\noindent a uniform distribution on the range \begin{equation}
  \left(\max\left(\max_{n \in [N]:\, \alpha_{nk} = l - 1} \left(\widetilde{\alpha^{\ast}}_{nk}\right), \widetilde{\gamma}_{k,l-1}\right), \min\left(\min_{n \in [N]:\, \alpha_{nk} = l} \left(\widetilde{\alpha^{\ast}}_{nk}\right), \widetilde{\gamma}_{k,l+1}\right)\right)
\end{equation}

In the above expression if there are no subjects whose latent attribute for dimension \(k\) is level \(l - 1\), the left-hand bound evaluates to \(\widetilde{\gamma}_{k,l-1}\). Similarly, if there are no subjects whose latent attribute for dimension \(k\) is level \(l\), the right-hand bound evaluates to \(\widetilde{\gamma}_{k,l+1}\).

For \(l = L - 1\), the full conditional is \begin{align}\label{bg4part2}
  & p(\widetilde{\gamma}_{kl} \mid \widetilde{\gamma}_{k,l-1}, \widetilde{\alpha^\ast}) \notag\\
  & = c_1 \cdot (\widetilde{\text{part2}}) \cdot (\widetilde{\text{part3}}) \notag\\
  & = c_2 \cdot \left[\prod_{n \in [N]: \alpha_{nk} = l-1} I\left(\widetilde{\alpha^\ast}_{nk} \in \Big(\widetilde{\gamma}_{k, l-1}, \widetilde{\gamma}_{kl}\Big]\right)\right] \cdot \left[\prod_{n \in [N]: \alpha_{nk} = l} I\left(\widetilde{\alpha^\ast}_{nk} \in \Big(\widetilde{\gamma}_{kl}, \widetilde{\gamma}_{k,l+1}\Big]\right)\right] \notag\\
  & \quad\quad \cdot \exp\left[-a (\widetilde{\gamma}_{kl} - \widetilde{\gamma}_{k, l-1})\right] \cdot I\left(\widetilde{\gamma}_{kl} \in (\widetilde{\gamma}_{k, l-1}, \infty)\right) \notag\\
  & = c_2 \cdot \left[\prod_{n \in [N]:\, \alpha_{nk} = l - 1} I(\widetilde{\gamma}_{kl} \geq \widetilde{\alpha^{\ast}}_{nk})\right] \cdot \left[\prod_{n \in [N]:\, \alpha_{nk} = l} I(\widetilde{\gamma}_{kl} < \widetilde{\alpha^{\ast}}_{nk})\right] \notag\\
  & \quad\quad \cdot \exp\left[-a (\widetilde{\gamma}_{kl} - \widetilde{\gamma}_{k, l-1})\right] \cdot I\left(\widetilde{\gamma}_{kl} \in (\widetilde{\gamma}_{k, l-1}, \infty)\right) \notag\\
  & = c_3 \cdot I\left(\widetilde{\gamma}_{kl} \geq \max\left(\max_{n \in [N]:\, \alpha_{nk} = l - 1} \left(\widetilde{\alpha^{\ast}}_{nk}\right), \widetilde{\gamma}_{k, l-1}\right)\right) \notag\\
      & \quad\quad \cdot I\left(\widetilde{\gamma}_{kl} < \min\left(\min_{n \in [N]:\, \alpha_{nk} = l} \left(\widetilde{\alpha^{\ast}}_{nk}\right), \infty\right)\right) \cdot \exp\left[-a (\widetilde{\gamma}_{kl} - \widetilde{\gamma}_{k, l-1})\right]
\end{align}

\noindent where if there are no respondents' whose latent state is at the highest level for dimension \(k\) (namely, \(L - 1\)), the right-hand bound evaluates to \(\infty\). We observe that the above density is that of a truncated exponential with one of four truncation ranges: \begin{enumerate}
\item \((\max_{n \in [N]:\, \alpha_{nk} = l - 1}(\widetilde{\alpha^{\ast}}_{nk}), \min_{n \in [N]:\, \alpha_{nk} = l}(\widetilde{\alpha^{\ast}}_{nk}))\)
\item \((\max_{n \in [N]:\, \alpha_{nk} = l - 1}(\widetilde{\alpha^{\ast}}_{nk}), \infty)\)
\item \((\widetilde{\gamma}_{k, l-1}, \min_{n \in [N]:\, \alpha_{nk} = l}(\widetilde{\alpha^{\ast}}_{nk}))\)
\item \((\widetilde{\gamma}_{k, l-1}, \infty)\)
\end{enumerate}

\subsection*{Covariance matrix and slope parameter for covariates}

We sample \((\widetilde{\lambda}, \Sigma)\) as in \citet[Section 2.8]{rossi2005bayesian}, by first sampling \(\Sigma\) with a Gibbs step collapsed on \(\widetilde{\lambda}\), and then using that value to sample \(\widetilde{\lambda}\) from its full conditional. To find the density for that first step, we integrate the full conditional of \((\widetilde{\lambda}, \Sigma)\) with respect to \(\widetilde{\lambda}\); that full conditional is proportional to \begin{align}
  p(\widetilde{\lambda}, \Sigma \mid \widetilde{\alpha^\ast}) & = c_1 \cdot (\widetilde{\text{part4}}) \cdot (\widetilde{\text{part5}}) \cdot (\widetilde{\text{part6}}) \notag\\
  & = c_2 \cdot (\det{\Sigma})^{-\frac{1}{2}(v_0 + K + 1)} \text{etr}\left(-\frac{1}{2}\Sigma^{-1}\right) \notag\\
  & \quad\quad \cdot (\det{\Sigma})^{-\frac{1}{2}N} \text{etr}\left\{-\frac{1}{2}\left(\widetilde{\alpha^\ast} - X \widetilde{\lambda}\right) \Sigma^{-1} \left(\widetilde{\alpha^\ast} - X \widetilde{\lambda}\right)^\prime\right\} \notag\\
  & \quad\quad \cdot (\det{\Sigma})^{-\frac{1}{2}D} (\det{I_D})^{-\frac{1}{2}K} \cdot \text{etr}\left\{-\frac{1}{2} \widetilde{\lambda} \Sigma^{-1} \widetilde{\lambda}^\prime\right\} \label{bg5part1}
\end{align}

Following \citet{rossi2005bayesian}, we let \begin{equation*}
Z = \begin{pmatrix}\widetilde{\alpha^\ast} \\ O\end{pmatrix}, \quad \Omega = \begin{pmatrix}X \\ I_D\end{pmatrix}
\end{equation*}

\noindent where \(O\) is of dimension \(D \times K\). Note that \begin{equation*}
\Omega^\prime \Omega = X^\prime X + I_D, \quad \Omega^\prime Z = X^\prime \widetilde{\alpha^\ast}.
\end{equation*}

\noindent Further, we define \(\widehat{L_1} = (X^\prime X)^{-1} X^\prime \widetilde{\alpha^\ast}\), the frequentist estimate of \(\widetilde{\lambda}\), and define \begin{equation*}
  \widehat{L_2} = (X^\prime X + I_D)^{-1} (X^\prime X \widehat{L_1}) = (X^\prime X + I_D)^{-1} X^\prime \widetilde{\alpha^\ast}
\end{equation*}

\noindent an estimate of \(\widetilde{\lambda}\) appropriate for this Bayesian context. We also let \begin{equation*}
S = (\widetilde{\alpha^\ast} - X \widehat{L_2})^\prime (\widetilde{\alpha^\ast} - X \widehat{L_2}) + \widehat{L_2}^\prime I_D \widehat{L_2}.
\end{equation*}

\noindent Following \cite{rossi2005bayesian}, \begin{align}
& (\widetilde{\alpha^\ast} - X \widetilde{\lambda})^\prime (\widetilde{\alpha^\ast} - X \widetilde{\lambda}) \nonumber\\
& = (Z - \Omega\widetilde{\lambda})^\prime (Z - \Omega\widetilde{\lambda}) - \widetilde{\lambda}^\prime I_D \widetilde{\lambda} \nonumber\\
& = (Z - \Omega\widehat{L_2})^\prime (Z - \Omega\widehat{L_2}) + (\widetilde{\lambda} - \widehat{L_2})^\prime \Omega^\prime \Omega (\widetilde{\lambda} - \widehat{L_2}) - \widetilde{\lambda}^\prime I_D \widetilde{\lambda} \nonumber\\
& = (\widetilde{\alpha^\ast} - X \widehat{L_2})^\prime (\widetilde{\alpha^\ast} - X \widehat{L_2}) + \widehat{L_2}^\prime I_D \widehat{L_2} + (\widetilde{\lambda} - \widehat{L_2})^\prime \Omega^\prime \Omega (\widetilde{\lambda} - \widehat{L_2})- \widetilde{\lambda}^\prime I_D \widetilde{\lambda} \nonumber\\
& = S + (\widetilde{\lambda} - \widehat{L_2})^\prime \Omega^\prime \Omega (\widetilde{\lambda} - \widehat{L_2}) - \widetilde{\lambda}^\prime I_D \widetilde{\lambda}
\end{align}

\noindent Using this, we have \begin{align}
  \eqref{bg5part1} & = c_2 \cdot (\det{\Sigma})^{-([v_0 + N] + K + 1)/2} \cdot \text{etr}\left\{-\frac{1}{2}(I_K + S) \Sigma^{-1}\right\} \notag\\
  & \quad\quad \cdot (\det{\Sigma})^{-D/2} \text{etr}\left\{-\frac{1}{2} (\widetilde{\lambda} - \widehat{L_2})^\prime \Omega^\prime \Omega (\widetilde{\lambda} - \widehat{L_2}) \Sigma^{-1}\right\} \notag\\
  & \quad\quad \cdot \text{etr}\left\{-\frac{1}{2} \widetilde{\lambda} \Sigma^{-1} \widetilde{\lambda}^\prime\right\} \cdot \text{etr}\left\{\frac{1}{2} \widetilde{\lambda} \Sigma^{-1} \widetilde{\lambda}^\prime\right\} \notag\\
  & = c_2 \cdot (\det{\Sigma})^{-([K + 1 + N] + K + 1)/2} \cdot \text{etr}\left\{-\frac{1}{2}(I_K + S) \Sigma^{-1}\right\} \notag\\
  & \quad\quad \cdot (\det{\Sigma})^{-D/2} \text{etr}\left\{-\frac{1}{2} \Omega^\prime \Omega (\widetilde{\lambda} - \widehat{L_2}) \Sigma^{-1} (\widetilde{\lambda} - \widehat{L_2})^\prime \right\}
\end{align}

\noindent the product of two terms, the first of which is the density of an inverse Wishart with variable \(\Sigma\), matrix parameter \(I_K + S\), and scalar parameter \(K + 1 + N\), and the second of which is the density of a matrix variate normal with variable \(\widetilde{\lambda}\), mean \(\widehat{L_2}\), and covariance \((\Omega^\prime \Omega)^{-1} \otimes \Sigma\) (see Supplementary Material C, also \citet{gupta1999matrix}). In the above we used the fact that \(v_0 = K + 1\).

Therefore \begin{align}
  p(\Sigma \mid \widetilde{\alpha^\ast}) & = \int p(\widetilde{\lambda}, \Sigma \mid \widetilde{\alpha^\ast}) d\widetilde{\lambda} \notag\\
  & = c_3 \cdot (\det{\Sigma})^{-([K + 1 + N] + K + 1)/2} \cdot \text{etr}\left\{-\frac{1}{2}(I_K + S) \Sigma^{-1}\right\}
\end{align}

an inverse Wishart with matrix parameter \(I_K + S\) and scalar parameter \(K + 1 + N\).

The full conditional of \(\widetilde{\lambda}\) is, using the above algebraic manipulations and proportionality, \begin{equation}
  \widetilde{\lambda} \mid \Sigma, \widetilde{\alpha^\ast} \sim N_{D, K}(\widetilde{\lambda}; \widehat{L_2}, (X^\prime X + I_D)^{-1} \otimes \Sigma)
\end{equation}

\subsection*{Sparsity matrix related parameter}

We sample \(\omega\) from its full conditional, which is \begin{align}
  p(\omega \mid \delta) & = c_1 \cdot (\text{part1}) \notag\\
  & = c_1 \cdot \left[\prod_{j=1}^J \prod_{h=1}^H \omega^{\delta_{hj}} (1 - \omega)^{1 - \delta_{hj}}\right] \cdot \omega^{\omega_0 - 1} (1 - \omega)^{\omega_1 - 1} \notag\\
  & = c_1 \cdot \omega^{(\sum_{j \in [J], h \in [H]} \delta_{hj} + \omega_0) - 1} (1 - \omega)^{(HJ - \sum_{j \in [J], h \in [H]} \delta_{hj} + \omega_1) - 1}
\end{align}

\noindent so \(\omega \mid \delta \sim \text{Beta}\left(\sum_{j \in [J], h \in [H]} \delta_{hj} + \omega_0,\, HJ - \sum_{j \in [J], h \in [H]} \delta_{hj} + \omega_1\right)\)

\newpage

\section*{Supplementary Material C}

Supplementary Material C contains information on the matrix variate normal distribution.

Quoting almost verbatim from \citet[page 55]{gupta1999matrix}:

\vspace{\baselineskip}

\noindent \textbf{Definition 2.2.1} (Gupta and Nagar): The random matrix \(X (p \times n)\) is said to have a matrix variate normal distribution with mean matrix \(M (p \times n)\) and covariance matrix \(\Sigma \otimes \Psi\) where \(\Sigma (p \times p) > 0\) and \(\Psi (n \times n) > 0\), if \(\text{vec}(X^\prime) \sim N_{pn}(\text{vec}(M^\prime), \Sigma \otimes \Psi)\), a multivariate normal distribution.

\vspace{\baselineskip}

In this manuscript and in these Supplementary Materials, to denote a matrix variate normal random variable \(X\) with mean \(M\) and covariance \(\Sigma \otimes \Psi\) we use the notation \(X \sim N_{p,n}(M, \Sigma \otimes \Psi)\).

\vspace{\baselineskip}

\noindent \textbf{Theorem 2.2.1} (Gupta and Nagar): If \(X \sim N_{p,n}(M, \Sigma \otimes \Psi)\), then the p.d.f. of \(X\) is given by \begin{equation*}
  (2\pi)^{-\frac{1}{2}np} (\det(\Sigma))^{-\frac{1}{2}n} (\det(\Psi))^{-\frac{1}{2}p} \text{etr}\left\{-\frac{1}{2} \Sigma^{-1} (X - M) \Psi^{-1} (X - M)^\prime\right\}
\end{equation*}

\noindent where of course \(X \in \mathbb{R}^{p \times n}\) and \(M \in \mathbb{R}^{p \times n}\).

\vspace{\baselineskip}

An important property of the matrix variate normal is the following:

\vspace{\baselineskip}

\noindent \textbf{Theorem 2.3.1} (Gupta and Nagar): If \(X \sim N_{p,n}(M, \Sigma \otimes \Psi)\), then \(X^\prime \sim N_{n,p}(M^\prime, \Psi \otimes \Sigma)\)

\newpage

\section*{Supplementary Material D}

Supplementary Material D contains alt text for Figure 1 of the manuscript.

The figure is a directed acyclic graph. There are two plates, each of which contains a subset of the vertices of the graph: \begin{itemize}
\item Plate \(j\) contains: \(\delta_j\), \(\beta_j\), \(\kappa_j\), \(Y_{nj}^\ast\), \(Y_{nj}\)
\item  Plate \(n\) contains: \(\alpha_{n}^\ast\), \(\alpha_{n}\), \(Y_{nj}^\ast\), \(Y_{nj}\)
\end{itemize}

In addition to the above vertices, there are five vertices not contained in any plate: \(\omega\), \(R\), \(\lambda\), \(V\), and \(\gamma\).

The directed edges between vertices are described in the following list, where for example the list item "\(a\) to \(b\)" indicates a directed edge from vertex \(a\) to vertex \(b\):

\begin{itemize}
\item \(\omega\) to \(\delta_j\)
\item \(\delta_j\) to \(\beta_j\)
\item \(\beta_j\) to \(Y_{nj}^\ast\)
\item \(\kappa_j\) to \(Y_{nj}\)
\item \(Y_{nj}^\ast\) to \(Y_{nj}\)
\item \(R\) to \(\lambda\)
\item \(R\) to \(V\)
\item \(V\) to \(\gamma\)
\item \(R\) to \(\alpha_{n}^\ast\)
\item \(\lambda\) to \(\alpha_{n}^\ast\)
\item \(\alpha_{n}^\ast\) to \(\alpha_{n}\)
\item \(\gamma\) to \(\alpha_{n}\)
\item \(\alpha_{n}\) to \(Y_{nj}^\ast\)
\end{itemize}
  
\end{document}